%% file: paper.tex
\documentclass[letterpaper,11pt]{article}
\usepackage[left=2.5cm,right=2.5cm,top=2.75cm,bottom=2.5cm,nohead]{geometry}

\usepackage{amsgen,amsbsy,amsopn,amssymb,color,graphicx,hyperref,endnotes}

\usepackage{mathptmx}
\usepackage{courier}
\usepackage[scaled=.90]{helvet}

\input{./tex/arxiv2print.tex}

\title{Amanuensis: The Programmer's Apprentice}

\date{}

\author{Thomas Dean$^{1,2}$\\
Maurice Chiang$^{2}$\\
Marcus Gomez$^{2}$\\
Nate Gruver$^{2}$\\
Yousef Hindy$^{2}$\\
Michelle Lam$^{2}$\\
Peter Lu$^{2}$\\
Sophia Sanchez$^{2}$\\
Rohun Saxena$^{2}$\\
Michael Smith$^{2}$\\
Lucy Wang$^{2}$\\
Catherine Wong$^{2}$}\footnotetext{Affiliations: $^{1}$Google Research, $^{2}$Stanford University}

\begin{document}


\input{./inputs/00_Title_Abstract.tex}


\input{./inputs/01_Introduction.tex}


\input{./inputs/02_Foundations.tex}


\input{./inputs/03_Interactions.tex}


\input{./inputs/04_Production.tex}


\addcontentsline{toc}{section}{Bibliography}

\input{./paper.bbl}


\appendix


\input{./inputs/A_Architecture.tex}


\input{./inputs/B_Bootstrapped.tex}


\def\notesname{Footnotes}
\theendnotes


\end{document}

%% file: tex/arxiv2print.tex
\def\urlh#1#2{{\href{#1}{#2}}}

\def\emdash{---}
\def\etal{{\em{et al}}}



%% file: inputs/00_Title_Abstract.tex



\pagenumbering{gobble}

\begin{titlepage}

  \maketitle

  \begin{abstract}
    This document provides an overview of the material covered in a course taught at Stanford in the spring quarter of 2018. The course draws upon insight from cognitive and systems neuroscience to implement hybrid connectionist and symbolic reasoning systems that leverage and extend the state of the art in machine learning by integrating human and machine intelligence. As a concrete example we focus on digital assistants that learn from continuous dialog with an expert software engineer while providing initial value as powerful analytical, computational and mathematical savants. Over time these savants learn cognitive strategies (domain-relevant problem solving skills) and develop intuitions (heuristics and the experience necessary for applying them) by learning from their expert associates. By doing so these savants elevate their innate analytical skills allowing them to partner on an equal footing as versatile collaborators \emdash{} effectively serving as cognitive extensions and digital prostheses, thereby amplifying and emulating their human partner's conceptually-flexible thinking patterns and enabling improved access to and control over powerful computing resources.
  \end{abstract}

\end{titlepage}

\newpage

\pagenumbering{roman}

\tableofcontents

\newpage

\pagenumbering{arabic}


%% file: inputs/01_Introduction.tex

\section{Introduction: Programmer's Apprentice}


Suppose you could merely imagine a computation, and a digital prostheses, an extension of your biological brain, would turn it into code that instantly realizes what you had in mind. Imagine looking at an image, dataset or set of equations and wanting to analyze and explore its meaning as an artistic whim or part of a scientific investigation. I don't mean you would use an existing software suite to produce a standard visualization, but rather you would make use of an extensive repository of existing code to assemble a new program analogous to how a composer draws upon a repertoire of musical motifs, themes and styles to construct new works, and tantamount to having a talented musical amanuensis who, in addition to copying your scores, takes liberties with your prior work, making small alterations here and there and occasionally adding new works of its own invention, novel but consistent with your taste and sensibilities.

Perhaps the interaction would be wordless and you would express your objective by simply focusing your attention and guiding your imagination, the prostheses operating directly on patterns of activation arising in your primary sensory, proprioceptive and associative cortex that have become part of an extensive vocabulary that you now share with your personal digital amanuensis. Or perhaps it would involve a conversation conducted in subvocal, unarticulated speech in which you specify what it is you want to compute and your assistant asks questions to clarify your intention and the two of you share examples of input and output to ground your internal conversation in concrete terms. 

More than thirty years ago, Charles Rich and Richard Waters published an MIT AI Lab technical report~\cite{RichandWatersAIM-87} entitled {\it{The Programmer's Apprentice: A Research Overview}}. Whether they intended it or not, it would have been easy in those days for someone to misremember the title and inadvertently refer to it as "The Sorcerer's Apprentice" since computer programmers at the time were often characterized as wizards and most children were familiar with the Walt Disney movie {\it{Fantasia}}, featuring music written by Paul Dukas inspired by Goethe's poem of the same name. The Rich and Waters conception of an apprentice was certainly more prosaic than the idea described above, but they might have had trouble anticipating the amount of code available in open-source repositories and the considerable computational power we carry about on our persons or can access through the cloud.

In any case, you might find it easier to imagine describing programs in natural language and supplementing your descriptions with input-output pairs. The programs could be as simple as regular expressions or SQL queries or as complicated as designing powerful simulators and visualization algorithms. The point is that there is a set of use cases that are within our reach now and that set will grow as we improve our natural language understanding and machine learning tools. I simply maintain that the scope of applications within reach today is probably larger than you think and that our growing understanding of human cognition is helping to substantially broaden that scope and significantly improve the means by which we interact with computers in general and a new generation of digital prostheses in particular. Here are just a few of the implications that might follow from pursuing a very practical and actionable modern version of The Programmer's Apprentice:


{\bf{Develop systems that enable human-machine collaboration on challenging design problems including software engineering:}} The objective of this effort is to develop digital assistants that learn from continuous dialog with an expert software engineer while providing initial value as powerful analytical, computational and mathematical savants. Over time these savants learn cognitive strategies (domain-relevant problem solving skills) and develop intuitions (heuristics and the experience necessary for applying them) by learning from their expert associates. By doing so these savants elevate their innate analytical skills allowing them to partner on an equal footing as versatile collaborators \emdash{} effectively serving as cognitive extensions and digital prostheses, thereby amplifying and emulating their human partner's conceptually-flexible thinking patterns and enabling improved access to and control over powerful computing resources. 


{\bf{Leverage and extend the current state of the art in machine learning by integrating human and machine intelligence:}} Current methods for training neural networks typically require substantial amounts of carefully labeled and curated data. Moreover the environments in which many learning systems are expected to perform are partially observable and non-stationary. The distributions that govern the presentation of examples change over time requiring constant effort to collect new data and retrain. The ability to solicit and incorporate knowledge gleaned from new experience to modify subsequent expectations and adapt behavior is particularly important for systems such as digital assistants with whom we interact and routinely share experience. Effective planning and decision making rely on counterfactual reasoning in which we imagine future states in which propositions not currently true are accommodated or steps taken to make them true~\cite{HassabisandMaguireTiCS-07}. The ability for digital assistants to construct predictive models of other agents \emdash{} so-called theory-of-mind modeling \emdash{} is critically important for collaboration~\cite{RabinowitzetalCoRR-18}.


{\bf{Draw insight from cognitive and systems neuroscience to implement hybrid connectionist and symbolic reasoning systems:}} Many state-of-the-art machine learning systems now combine differentiable and non-differentiable computational models. The former consist of fully-differentiable connectionist artificial neural networks. They achieve their competence by leveraging a combination of distributed representations facilitating context-sensitive, noise-tolerant pattern-recognition and end-to-end training via backpropagation. The latter, non-differentiable models, excel at manipulating representations that exhibit combinatorial syntax and semantics, are said to be full systematic and compositional, and can directly and efficiently exploit the advantages of traditional von Neumann computing architectures. The differences between the two models are at the heart of the connectionist versus symbolic systems debate that dominated cognitive science in 80's and continues to this day~\cite{OReillyetalTACO-14,FodorandPylyshynCOGNITION-88}. Rather than simulate symbolic reasoning within connectionist models or vice a versa, we simply acknowledge their strengths and build systems that enable efficient integration of both types of reasoning.


{\bf{Take advantage of advances in natural language processing to implement systems capable of continuous focused dialog:}} Language is arguably the most important technical innovation in the history of humanity. Not only does it make possible our advanced social skills, but it allows us to pass knowledge from one generation to the next and provides the foundation for mathematical and logical reasoning. Natural language is our native programming language. It is the way we communicate plans and coordinate their execution. In terms of expressiveness, it surpasses modern computer programming languages, but its capability for communicating imprecisely and even incoherently, and our tendency for utilizing that capability makes it a poor tool for programming conventional computers. That said it serves us well in training scientists and engineers to develop and apply more precise languages, and its expressiveness along with our facility using it make it an ideal means for humans and AI systems to collaborate. The consolidation and subsequent recall and management of episodic memory is a key part of what makes us human and enables our diverse social behaviors. Episodic memory makes it possible to create and maintain long-term relationships and collaborations~\cite{PritzeletalICML-17,MoscovitchetalARP-16,OReillyetalCS-15}.


{\bf{Think seriously about how such technology might ultimately be employed to build brain-computer-interfaced prostheses:}} This exercise primarily relies on the use of natural language to facilitate communication between the expert programmer and apprentice AI system. The AI system learns to use natural language in much the same way as a human apprentice would \emdash{} as a flexible and expressive tool to capture and convey understanding and recognize and resolve misunderstanding and ambiguity. The AI system interacts with computing hardware through a highly instrumented integrated development environment. Essentially, the AI system can read, write, execute and debug code by simply thinking \emdash{} reading and writing to a differentiable neural computing interface~\cite{GravesetalNATURE-16}. It can also directly sense code running by reading from {\tt{STDERR}} and {\tt{STDIO}}, parsing output from the debugger and collecting and analyzing program traces. The same principles could be applied to develop digital prostheses employed for a wide range of intelligence-enhancing human-computer interfaces.


\subsection{Resources}


This document attempts to optimize for the student or software engineer knowledgeable about neural networks and interested primarily in understanding how one might go about building a system along the lines of the programmer's apprentice. It is my experience that this audience has relatively little appetite for details about relevant work in cognitive and systems neuroscience that has informed the design sketched in these pages. The course website for the class I taught at Stanford in the 2018 Spring quarter serves as an extensive resource for those reading this document. It includes all of the {\urlh{https://web.stanford.edu/class/cs379c/calendar.html}{lectures}} and {\urlh{https://web.stanford.edu/class/cs379c/class_messages_listing/index.html}{discussion notes}} for the class and I refer to it often in this document as a source of supplementary information.


%% file: inputs/02_Foundations.tex

\section{Foundation: Cognitive Neuroscience}


This document is not intended to provide the reader with a short course in cognitive science, artificial intelligence, natural language processing, machine learning, artificial neural networks, or automated code synthesis / automatic inductive programming, and is certainly not intended to cover all these disciplines in any but the most cursory of detail. The primary goal is to explore the possibility of building digital assistants that considerably extend our ability to solve complex engineering problems with a emphasis here on software engineering. A secondary goal is to explain how the field of neuroscience is helping to achieve our primary goal.

The fields of cognitive and systems neuroscience are playing an important role in directing and accelerating research on artificial neural network systems. Much of this work predates and helped give rise to the especially exciting work on connectionist models in the 1980s. However, in the nearly 40 intervening years, a great deal of progress has been made, much of it due to improved methods for studying the behavior of awake behaving animal subjects and human beings in particular. Indeed, this work is undergoing a renaissance fueled by even more powerful methods for observing brain activity in human beings in the midst of solving complex cognitive tasks.

The field of automatic programming, after decades of steady, often quite practical research on using symbolic methods \emdash{} much of it originating in labs outside the United States, is seeing a renewed interest in artificial neural networks. It remains to be seen whether artificial neural networks will have a significant impact on code synthesis, however there appear to be opportunities to leverage what we know about both natural and artificial neural networks to make progress, and hybrid systems that combine both connectionist and traditional symbolic methods may have the best chance of pushing the state-of-the-art significantly beyond its present level.


\subsection{Memory}


We begin with the problem of how to represent information in memory. In the case of the programmer's apprentice, relevant information includes the type of items that software engineers routinely think about in plying their trade such as algorithms, data structures, interfaces, programs, subroutines and tools such as assemblers, compilers, debuggers, interpreters, parsers and syntax checkers. Then there are the things that programmers generally do not think about explicitly but that concern how they solve problems and organize their thoughts, including, for example, the design strategies we learn in computer science courses such as divide-and-conquer, dynamic-programming and recursion. Finally, there is strategic organizational information of a sort that plays a role in any complex individual or collaborative effort including plans, tasks, subtasks, specifications and requirements.

All of this information has to be encoded in memory and made accessible when required to perform cognitive tasks. Information, whether in a computer or a brain, tends to move around depending on what is to be done with it, and, at least in biological brains, it is constantly changing. In biological brains, it is difficult if not impossible to think about something without changing it. In building systems inspired by biological brains we have somewhat more control over such changes, but control comes at a cost. We make no distinction between concrete and abstract thoughts \emdash{} all thoughts are abstract whether they represent atoms or bits. We will on occasion refer to memories as being short- or long-term but the distinction doesn't begin to address real issues. When we talk about episodic memory, it may seem that we are referring to some sort of permanent or archival memory, but that's not the case.

Since this document is more condensed precis than unabridged thesis, we need some way of navigating the huge space of ideas relating to biological and artificial brains as they pertain to building digital assistants and automatic programming. I'll begin by pointing out that language, programs and plans are all usefully thought of as having hierarchical, recursive structure. It also makes sense to think of brains as being organized as such~\cite{Ballard2015,Kurzweil2012,DeanAMAI-06,GeorgeandHawkinsIJCNN-05,DeanAAAI-05,Hawkins04}, and the human brain apparently employs hierarchical models to make sense of the world in which it evolved. 

To the untutored mind, the world is essentially flat. We impose hierarchical structure to make understanding it more tractable. We ingest sequences of observations as input and execute sequences of actions as output. What goes on between is complicated. Rather than immediately focusing on how biological and artificial brains learn and apply hierarchical models, we start by considering the simpler problem of how we might represent a {\it{subroutine}}, the smallest fungible unit of activity for our purposes. Subroutines can be used to kick a soccer ball or implement simple program transformations in a neural-network architecture.


The following assumes familiarity with artificial neural networks. We begin with the simplifying assumption that subroutines can be represented as tuples consisting of a set of operands represented as high-dimensional embedding vectors, a weight matrix representing the transformation and a product vector space in which to embed the result. In applying this idea to program transformations, assume that each operand corresponds to the embedding of an abstract-syntax-tree representation of a code fragment, w.l.o.g., any non-terminal node in the AST of a syntactically well-formed program. In the remainder of this section and the next, we use the following abstractions and abbreviations:
\begin{itemize}
\item {\it{prefrontal cortex}} (PFC) including attention, conscious access, reward-based-learning and executive control~\cite{WangetalNATURE-NEUROSCIENCE-18,KrieteetalPNAS-13};
\item {\it{entorhinal-hippocampal complex}} (EHC) in its role as primary interface between the hippocampus and neocortex~\cite{OReillyetalCS-15,OReillySCIENCE-06};
\item {\it{global workspace}} (GW) broadly distributed cortical circuits connected through long-range excitatory axons~\cite{DehaeneetalPNAS-98,Baars1988};
\item {\it{basal ganglia}} (BG) for its role in action selection and dynamic gating to direct input to the prefrontal cortex\cite{OReillyetalLEABRA-16,KrieteetalPNAS-13};
\item {\it{semantic memory system}} (SMS) including areas of the brain responsible for mathematical and abstract thought~\cite{Tulving1972,BinderandDesaiTiCS-11};
\item {\it{episodic memory system}} (EMS) including episodic memory management and memory-based parameter adaptation~\cite{SprechmannetalICLR-18,PritzeletalICML-17};
\item {\it{differentiable neural computer}} (DNC) as the interface to the integrated development environment prostheses~\cite{GravesetalNATURE-16,GravesetalCoRR-14};
\item {\it{abstract syntax-tree}} (AST) is a representation of the abstract syntactic structure of a source-code program~\cite{DevlinetalICLR-18,WangetalCoRR-17};
\end{itemize}

The second introductory lecture ({\urlh{https://web.stanford.edu/class/cs379c/calendar_invited_talks/lectures/04/05/slides/index.html}{HTML}}) for the Stanford {\urlh{https://web.stanford.edu/class/cs379c/}{course}} associated with this document provides a high-level overview of the relevant research in cognitive and systems neuroscience. Annotations of the form "Slide \#" link to relevant slides in the introductory lecture. For example {{\urlh{https://web.stanford.edu/class/cs379c/calendar_invited_talks/lectures/04/05/slides/index.html\#CS379C_INTRODUCTORY_LECTURE_2_SLIDE_02}{Slide~2}}} covers the key anatomical landmarks in the human brain mentioned in this document. While the material in the remainder of this section refers to work in cognitive and systems neuroscience, the discussion here emphasizes applications of what we've learned from neuroscience, and so the interested reader is encouraged to at least skim the above-linked lecture notes.


Referring to the abstractions and abbreviations introduced in the previous section, reading a program from {\tt{STDIO}} \emdash{} the analog of a human programmer reading a program displayed on a monitor \emdash{} will result in \emdash{} at least \emdash{} two different internal representations of the resulting AST: an embedding vector in the SMS and a key-value representation in the DNC. The former allows us manipulate programs and program fragments as fully-differentiable representations within distributed models. The latter allows us to modify, execute and share code in a human-accessible format, fully compatible with our software-development toolchain.

Following~\cite{PritzeletalICML-17}, we assume EMS consists of initial-state-action-reward-next-state tuples of the form $(s_{t},\;a_{t},\;r_{t},\;s_{t+1})$. State representations $s_{t}$ have to be detailed enough to reconstruct the context in which the action is performed and yet concise enough to be practical. Suppose the PFC directs the activation of selected circuits in the SMS via the global workspace (GW) \emdash{} see Slides~{{\urlh{https://web.stanford.edu/class/cs379c/calendar_invited_talks/lectures/04/05/slides/index.html\#CS379C_INTRODUCTORY_LECTURE_2_SLIDE_03}{3}}} and~{{\urlh{https://web.stanford.edu/class/cs379c/calendar_invited_talks/lectures/04/05/slides/index.html\#CS379C_INTRODUCTORY_LECTURE_2_SLIDE_05}{5}}} \emdash{} in accord with Dehaene~\etal{}~\cite{DehaeneetalSCIENCE-17,Dehaene2014} assuming a prior that generates low-dimensional thought vectors~\cite{BengioCoRR-17}. The state representation $s_{t}$ encodes the attentional state that served to identify representations in SMS relevant to $a_{t}$ allowing the EHC to produce the resulting state $s_{t+1}$. Given $s_{t}$ we can reproduce the activity recorded in the EMS, and, in principle, incorporate multiple steps and contingencies in a policy constituting a specialized program-synthesis or program-repair subroutine.

Such subroutines would include repairing a program in which a variable is introduced but not initialized, or when it is initialized but ambiguously typed or scoped. As another example, a variable is initialized as {\tt{VOID}} and subsequently assigned an integer value in some but not all branches of a conditional statement. Other examples of repair routines include problems with the use of comparison operators, e.g., two conditional branches with the same inequality, the {\tt{is}} operator is used instead of {\tt{is not}}, or vice versa, confusion involving {\tt{A is not None}}, {\tt{A not None}} and {\tt{A != None}}, and problems involving class methods, e.g., when {\tt{self}} accessor is missing from a variable, e.g., {\tt{mode = 'manual'}} instead of {\tt{self.mode = 'manual'}}~\cite{ShinetalICLR-18b,DevlinetalICLR-18,WangetalCoRR-17}.

Attentional machinery in the prefrontal cortex (PFC) populates the (GW) by activating circuits relevant to the current input and internal state, including that of the DNC and any ongoing activity in (SMS) circuits produced by previous top-down attention and bottom-up sensory processing. The PFC in its role as executive arbiter identifies operators in the form of policy subroutines and then enlists the EHC to \emdash{} using terminology adapted from Von Neumann machines \emdash{} to load registers in short-term memory and perform operations by using fast weights to transform the contents of the loaded registers into product representations that can either be fed to associative embeddings, temporarily stored in other registers or used to modify the contents of the DNC thereby altering the AST representation of the target code and updating the display to provide feedback to the human programmer.

The primate cortex appears to be tiled with columnar structures referred to as {\it{cortical columns}}. Some neuroscientists believe that all of these columns compute the same basic function. However, there is considerable variation in cell type, thickness of the cortical layers, and the size of the dendritic arbors to question this hypothesis. The prefrontal cortex is populated with a type of neuron, called a {\it{spindle neuron}}, similar in some respects to the {\it{pyramidal cells}} found throughout the cortex, that allow rapid communication across the large brains of great apes, elephants, and cetaceans. Although rare in comparison to other neurons, spindle neurons are abundant and quite large in humans and apparently play an important role in consciousness and attentional networks \emdash{} see {{\urlh{https://web.stanford.edu/class/cs379c/calendar_invited_talks/lectures/04/05/slides/index.html\#CS379C_INTRODUCTORY_LECTURE_2_SLIDE_04}{Slide~4}}}.

The corresponding artificial neural network architecture for the programmer's apprentice application consists of a hierarchy of specialized networks with a relatively dense collection of feedforward and feedback connections that enable recurrent state, attentional focus and the management of specialized memory systems that persists across different temporal scales \emdash{} see {{\urlh{https://web.stanford.edu/class/cs379c/calendar_invited_talks/lectures/04/05/slides/index.html\#CS379C_INTRODUCTORY_LECTURE_2_SLIDE_06}{Slide~6}}}. Individual networks are specialized to serve different types of representation, employing convolutional networks, gated-feedback recurrent networks and specialized embedding models. All of these networks are distributed representations that encode information in high-dimensional vector spaces such that different dimensions can be trained to represent different features allowing attentional mechanisms to emphasize or modify encodings so as to alter their meaning.

These attentional networks are connected to regions throughout the cortex and are trained via reinforcement learning to recognize events worth attending to according to the learned value function. Using extensive networks of connections \emdash{} both incoming and outgoing, attentional networks are able to create a composite representation of the current situation that can serve a wide range of executive cognitive functions, including decision making and imagining possible futures. The basic idea of a neural network trained to attend to relevant parts of the input is key to a number of the systems that we'll be looking at.

To understand attentional networks, think about an encoder-decoder network for machine translation. As the encoder digests each word in the sequence of words that constitute the input sentence, it produces a representation \emdash{} Geoff Hinton refers to these as {\it{thought clouds}} in analogy to the iconic clouds that you see in comic strips \emdash{} of the sentence fragment or {\it{prefix}} that it has seen so far. Because the sentence is ingested one word at a time \emdash{} generally proceeding from left to right \emdash{} the resulting thought cloud will tend to emphasize the meaning of the most recently ingested words in each prefix. You could encode the entire input sentence and then pass the resulting representation on to the decoder, but earlier words in the sentence will receive less attention that later words. Alternatively, you could introduce a new network layer that takes as input encodings of all the sentence prefixes seen so far and trains the new layer \emdash{} thereby taking advantage of the power of gradient descent \emdash{} to produce a composite representation that emphasizes those parts of the input that are most relevant in decoding / generating the next word in the output.

The programmer's apprentice is implemented as an instance of an hierarchical neural network architecture. It has a variety of conventional inputs that include speech and vision, as well as output modalities including speech and text. In these respects, it operates like most existing commercial personal assistants \emdash{} see {{\urlh{https://web.stanford.edu/class/cs379c/calendar_invited_talks/lectures/04/05/slides/index.html\#CS379C_INTRODUCTORY_LECTURE_2_SLIDE_07}{Slide~7}}}. It differs substantially, however, in terms of the way in which the apprentice interacts with the programmer. It is useful to think of the programmer and apprentice as pair programming, with the caveat that the programmer is in charge, knows more than the apprentice does \emdash{} at least initially, and is invested in training the apprentice to become a competent software engineer. One aspect of their joint attention is manifest in the fact that they share a browser window. The programmer interacts with the browser in a conventional manner while the apprentice interacts with it as though it is part of its body directly reading and manipulating the HTML using the browser API. The browser serves both programmer and apprentice as an encyclopedic source of useful knowledge as well as another mode of interaction and teaching.

The spatial relationships among the ganglion cells in the retina are preserved in the activity of neurons found in the primary visual \emdash{} or {\it{striate}} \emdash{} cortex. Most sensory and motor areas maintain similar modality-specific topographic relationships. Shown {{\urlh{https://web.stanford.edu/class/cs379c/calendar_invited_talks/lectures/04/05/slides/index.html\#CS379C_INTRODUCTORY_LECTURE_2_SLIDE_10}{here}}}, for example, are Wilder Penfield's famous motor and somatosensory {\urlh{https://en.wikipedia.org/wiki/Cortical_homunculus}{homunculi}} depicting the areas and proportions of the human brain dedicated to processing motor and sensory functions. Scientists have observed that the area devoted to the hands tend to be larger among pianists, while the relevant areas in the brains of amputees typically become significantly smaller \emdash{} see {{\urlh{https://web.stanford.edu/class/cs379c/calendar_invited_talks/lectures/04/05/slides/index.html\#CS379C_INTRODUCTORY_LECTURE_2_SLIDE_10}{Slide~10}}}.

We imagine the programmer's apprentice with a body part consisting of an instrumented {\it{integrated development environment}} (IDE). Alternatively you might think of it as a prosthetic device. It is not, however, something that you can simply remove or replace with an alternative device outfitted with a different interface or supporting different functions and expect it to immediately respond to your attempts to control it \emdash{} it is not a plug-and-play device. Like the legs you were born with or the prosthesis replacing an amputee's severed arm, you have to learn how to use these devices. Architecturally, the apprentice's prosthetic IDE is an instance of a {\it{differentiable neural computer}} (DNC) introduced by Alex Graves and his colleagues at DeepMind. The assistant combined with its prosthetic IDE is neural network that can read from and write to an external memory matrix, combining the characteristics of a random-access memory and set of memory-mapped device drivers and programmable interrupt controllers. The interface supports a fixed number of commands and channels that provide feedback. You can think of it as roughly similar to an Atari game console \emdash{} see {{\urlh{https://web.stanford.edu/class/cs379c/calendar_invited_talks/lectures/04/05/slides/index.html\#CS379C_INTRODUCTORY_LECTURE_2_SLIDE_11}{Slide~11}}}.

What is left out of this account so far includes how we might take advantage of semantics in the form of executing code and examining traces in order to better understand the consequences of the changes just made. Presumably, wrapping a code fragment in a function and executing the function with different input to examine changes in the state variables could be used as a distal reinforcement signal providing intermediate rewards useful in debugging subroutines. As pointed out earlier, subroutines designed to modify code are likely to involve many conditional choices and so it is important for subroutine policies to be highly conditioned on the status of specific state variables. Indeed a technique such as model-based parameter adaptation may be perfectly suited to providing such context-sensitive adaptations.

Perhaps this next observation seems obvious, but it is worth keeping in mind that the human brain does a great deal of (parallel) processing that never rises to the level of conscious attention. The executive control systems in the prefrontal cortex don't have micromanage everything. Every thought corresponds to a pattern of activity in one or more neural circuits in the brain or beyond in the peripheral nervous system. One pattern of activity inevitably leads to another in the same or another set of neurons. For example, patterns of activity that begin in the sensory cortex can lead to patterns of activity in the motor cortex and can have consequences elsewhere in the brain, e.g., in the cerebellar cortex resulting in speech, or external to the central nervous system as in the case of neurons that propagate through the peripheral nervous system causing muscles to contract and extend thereby making your limbs and torso move. 

Every new observation, every act of creating a new thought or revisiting an old one produces even more activity in the brain resulting in new thoughts some of which are ignored as their reverberations weaken and die and others that spawn new thoughts and proliferate under the influence of reentrant production of activity and the active encouragement of conscious attention in a perpetually self reinforcing, reimagining and self-analyzing cycle of recurrent activity. Meta-reinforcement learning supports the sort of diverse activity one might expect from a system that selects activity to attend to and then makes available in the global workspace for ready access by other systems. Sustaining a collection of such activated circuits would help to provide a context, serve to maintain a stack of policies, guide switching between them, support caching partial results for later use, reconstructing necessary state as needed when restoring policy after a recursive descent.

When you think of building systems that can develop new algorithms it is instructive the think about the simple case of learning to sort lists from input-output pairs. The bubble sort algorithm is generally regarded as the easiest to come up with, but even then it is easier if you start with simple I/O pairs like $[A,\;B] \rightarrow{} [A,\;B], [B,\;A] \rightarrow{} [A,\;B]$ and work up to longer lists \emdash{} referred to as curriculum learning~\cite{BengioetalICML-09}. As Dan Abolafia pointed out in his class {\urlh{https://web.stanford.edu/class/cs379c/calendar_invited_talks/lectures/04/24/index.html}{presentation}}, it is relative easy to learn to sort lists of length no more than $n$, but substantially more difficult to learn an algorithm that works for lists of arbitrary length, without the ability to construct a simple inductive proof of correctness. Logic and basic theorem proving are certainly important in learning to write programs. You might want to look at the Coq proof assistant for a glimpse at the future of algorithm development~\cite{BertotandCasteranCOQ-04}.


\begin{figure}
  \begin{center}
    \includegraphics[width=\textwidth]{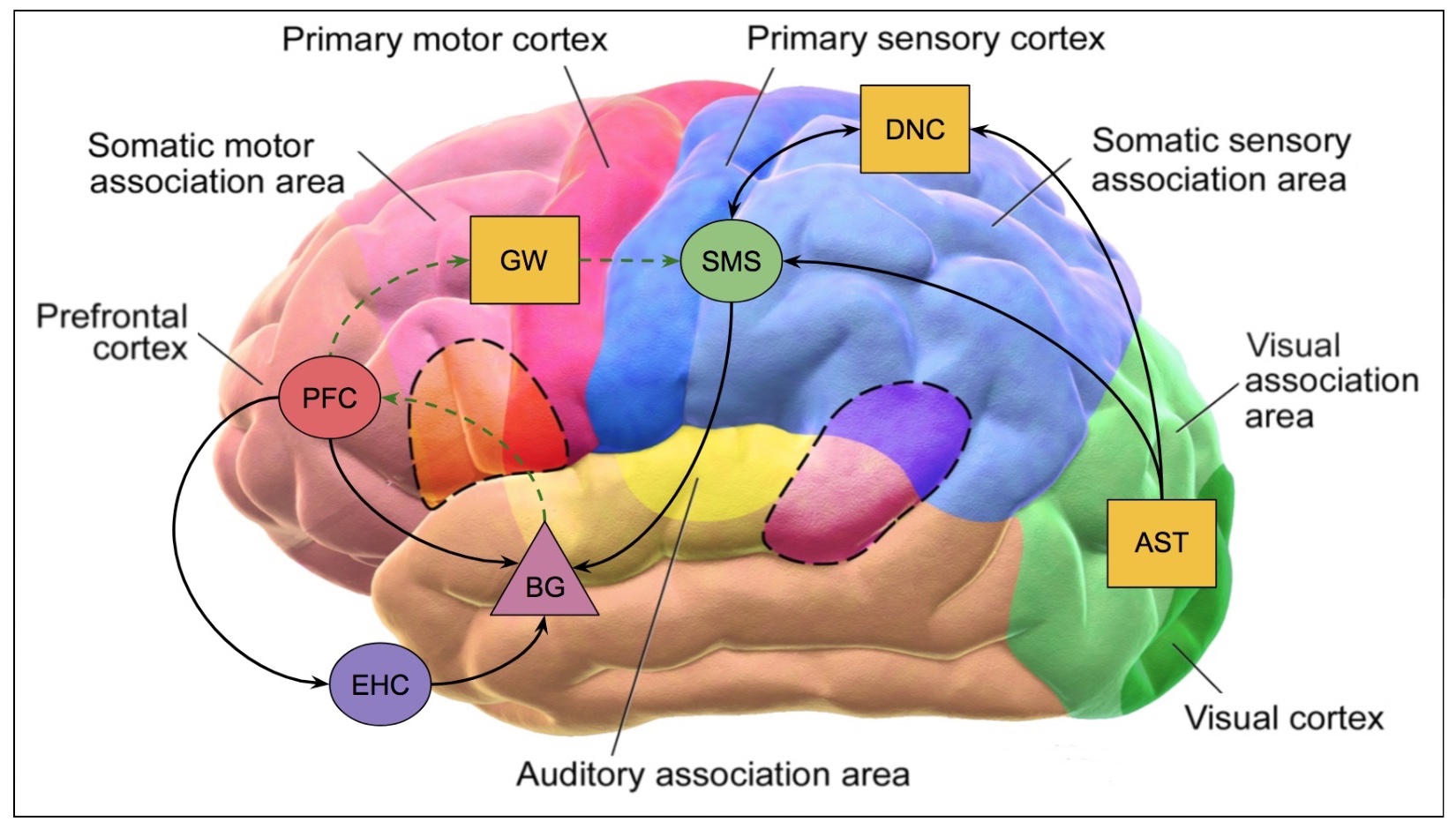}
  \end{center}
  \caption{This figure highlights the primary architectural components mentioned in the main text superimposed over an anatomical rendering of the human brain identifying related cortical and sub-cortical landmarks. The triangle and three ovals and match the shape and color conventions employed in O'Reilly~\cite{OReillySCIENCE-06} where you will find a substantially more detailed explanation of the underlying biological model. The three gold square shapes denote abstract architectural structures and not anatomical features. The acronyms are expanded and explained in the text.}
  \label{fig_Integrated_Architecture_Integrated_Figure}
\end{figure}


Figure~\ref{fig_Integrated_Architecture_Integrated_Figure} shows a diagram of the human brain overlaid with a simplified architectural drawing. The box shapes represent abstract systems and the oval and triangular shapes represent anatomical features for which we can supply computational models. For example, the box labeled GW represents the global workspace which performs a particular function in the architecture, but actually spans a good portion of the neocortex. Whereas the triangle labeled BG represents a group of subcortical nuclei called the basal ganglia situated at the base of the forebrain.

The box labeled AST represents a form of sensory input corresponding to the ingestion of abstract syntax trees representing code fragments. The oval labeled SMS represents semantic memory and the box labeled DNC corresponds to a differentiable neural computer. When the system ingests a new program fragment the resulting AST is encoded in the SMS as an embedding vector and simultaneously as a set of key-value pairs in the DNC. Here we think of the DNC as a body part or external prosthesis with corresponding maps in the somatosensory and motor cortex that enable reading and writing respectively \emdash{} see Slides~{{\urlh{https://web.stanford.edu/class/cs379c/calendar_invited_talks/lectures/04/05/slides/index.html\#CS379C_INTRODUCTORY_LECTURE_2_SLIDE_10}{10}}} and~{{\urlh{https://web.stanford.edu/class/cs379c/calendar_invited_talks/lectures/04/05/slides/index.html\#CS379C_INTRODUCTORY_LECTURE_2_SLIDE_11}{11}}} mentioned earlier.

Our explanation of the architecture proceeds top down, as it were, with a discussion of executive function in the prefrontal cortex. The GW provides two-way connection between structures in the prefrontal cortex and homologous structures of a roughly semantic character throughout the rest of neocortex thereby enabling the PFC to listen in on diverse circuits in the neocortex and select a subset of such circuits for attention. Stanislas Dehaene describes this process as one of the primary functions of consciousness, but we need not commit ourselves to such interpretation here.

Not only does the PFC selectively activate circuits but it can also maintain the activity such circuits indefinitely as constituents of working memory. Since this capability is limited by the capacity of the PFC, the content of working memory is limited and adding new constituents may curtail the activation of existing constituents. In practice, we intend to model this capability using meta-reinforcement learning~\cite{WangetalNATURE-NEUROSCIENCE-18} (MRL) in which the MRL system relies on the GW network to sample, evaluate and select constitutuent circuits guided by a suitable prior~\cite{BengioCoRR-17} and past experience and then maintain their activity by a combination of memory networks~\cite{WestonetalCoRR-14} and fast weights~\cite{BaetalCoRR-16}. 


Meta-reinforcement learning serves a second complementary role in the PFC related to executive function. We will refer to the first role as MRL-A for "attention" and the second as MRL-P for "planning". MRL-A is trained to focus attention on relevant new sensory input and new interpretations of and associations among prior perceptions and thoughts. MRL-P is trained to capitalize on and respond to opportunities made available by new and existing constituents in working memory. Essentially MRL-P is responsible for the scheduling and deployment of plans relevant to recognized opportunities to act. These plans are realized as policies trained by reinforcement learning from traces of past experience or constructed on the fly in response to unexpected / unfamiliar contingencies by recovering and reimagining past activities recovered from episodic memory \emdash{} see~{{\urlh{https://web.stanford.edu/class/cs379c/calendar_invited_talks/lectures/04/05/slides/index.html\#CS379C_INTRODUCTORY_LECTURE_2_SLIDE_13}{Slide~13}}}.

MRL-A and MRL-P could be implemented as a single policy, but it is simpler to think of them as two coupled systems, one responsible for focusing attention by constantly assessing changes in (neural) activity throughout the global workspace, and a second responsible for overseeing the execution of plans in responding to new opportunities to solve problems. MRL-A is as a relatively straightforward reinforcement learning system independently performing its task largely a function of whatever neural activity is going on in the GW, its attentional network and the prior baked into its reward function. MRL-P could be implemented along the lines of the Imagination-Augmented Agent (I2A) architecture~\cite{WeberetalCoRR-17} or the related Imagination-Based Optimization~\cite{HamricketalCoRR-17} and Imagination-Based Planning~\cite{PascanuetalCoRR-17} systems.


The remaining parts of the architecture involve the interplay between the PFC and the semantic and episodic memory systems as facilitated by the basal ganglia and hippocampus. If we had a policy pre-trained for every possible contingency, we would be nearly done \emdash{} let MRL-A draw attention to relevant internal and external activity and then design a simple just-in-time greedy scheduler that picks the policy with the highest reward given the state vector corresponding to the current content of working memory. Unfortunately, the life of an apprentice programmer is not nearly so simple. The apprentice might listen to advice from a human programmer or watch someone solve a novel coding problem or repair a buggy program. Alternatively, it may be relatively simple to adapt an existing policy to work in the present circumstances. However, making progress on harder problems will depend on expert feedback or having an existing reward function that generalizes to the problem at hand. 

The hippocampus is perhaps best known for its role in supporting spatial reasoning. A type of pyramidal neuron called a {\it{place cell}} has been shown to become active when an experimental animal enters an area of a maze that it has visited before. However, the hippocampus plays a much larger role in memory by representing not just the "where" of experience but also the "when". The manner in which we employ short- and long-term memory is very different. We might construct a representation of our current situation in short-term memory, drawing upon our long-term memory to provide detail. 

The two memory systems are said to be complementary in that they serve different purposes, one provides an archival record of the past while the other serves as a scratchpad for planning purposes. In retrieving a memory there is a danger that we corrupt the long-term memory in the process of subsequent planning. This isn't simply an academic question, it is at the heart of how we learn from the past and employ what we've learned to think about the future. Our subtle memory systems enable us to imagine solutions to problems that humans have never faced, and account for a good deal of our incredible adaptivity. In several lectures, we will explore architectures that support such flexibility \emdash{} see {{\urlh{https://web.stanford.edu/class/cs379c/calendar_invited_talks/lectures/04/05/slides/index.html\#CS379C_INTRODUCTORY_LECTURE_2_SLIDE_14}{Slide~14}}}.


\subsection{Actions}


The basal ganglia in cognitive models such as the one described by Randall O'Reilly's in his {\urlh{https://web.stanford.edu/class/cs379c/calendar_invited_talks/lectures/04/12/index.html}{presentation}} in class, play a central role in action selection. This seems like a good opportunity to review how actions are represented in deep-neural-network implementations of reinforcement learning. Returning to our default representation for the simplest sort of episodic memory, $(s_{t},\;a_{t},\;r_{t},\;s_{t+1})$, it's easy to think of a state $s$ as a vector $s \in{} \mathbb{R}^{n}$ and a reward $r$ as a scalar value, $r \in{} \mathbb{R}$, but how are actions represented?

Most approaches to deep reinforcement learning employ a tabular model of the policy implying a finite \emdash{} and generally rather small \emdash{} repertoire of actions. For example, most of the experiments described in Wayne~\etal{}~\cite{WayneetalCoRR-18} (MERLIN) six-dimensional one-hot binary vector that maps a set of six actions: move forward, move backward, rotate left with rotation rate of 30, rotate right with rotation rate of 30, move forward and turn left, move forward and turn right. The action space for the grid-world problems described in Rabinowitz~\etal{}~\cite{RabinowitzetalCoRR-18} (ToMnets) consists of four movement actions: up, down, left, right and stay \emdash{} see {{\urlh{https://web.stanford.edu/class/cs379c/calendar_invited_talks/lectures/04/05/slides/index.html\#CS379C_INTRODUCTORY_LECTURE_2_SLIDE_15}{Slide~15}}}.

The programmer's apprentice (PA) operates on programs represented as trees, where the set of actions includes basic operations for traversing and editing trees \emdash{} or more generally directed-graphs with cycles if you assume edges in abstract syntax trees corresponding to loops, recursion and nested procedure calls, i.e., features common to nearly all the programs we actually care about. We still have a finite number of actions since for any given project we can represent the code base as a directed-acyclic graph with annotations to accommodate procedure calls and recursion, and use attention to direct and contextualize a finite set of edit operations.

Pritzel~\etal{}~\cite{PritzeletalCoRR-17} employ a semi-tabular representation of an agent's experience of the environment possessing features of episodic memory including long-term memory, sequentiality and context-based lookups. The representation called a {\it{differential neural dictionary}} (DND) is related to Graves~\etal{}~\cite{GravesetalNATURE-16} DNC. The programmer's apprentice is better suited to Vinyals~\etal{}~\cite{VinyalsetalNIPS-15} related idea of a {\it{pointer-network}} designed to learn the conditional probability of an output sequence with elements that are discrete tokens corresponding to positions in an input sequence \emdash{} see related work in natural language processing by Merity~\etal{}~\cite{MerityetalCoRR-16} on {\it{pointer sentinels}}.


\setcounter{figure}{1}


\begin{figure}
  \begin{center}
    \includegraphics[width=5.0in]{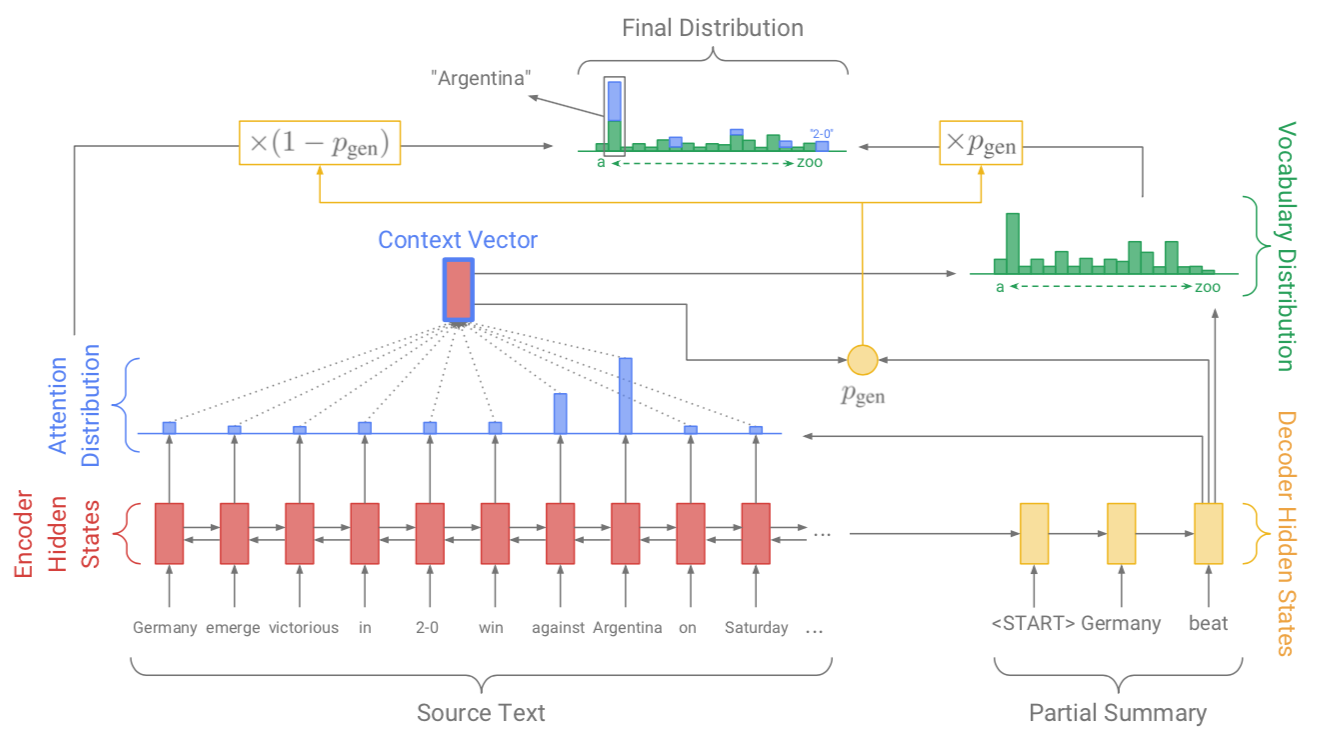}
  \end{center}
  \caption{The sequence-to-sequence encoder-decoder attentional model shown here uses a specialized memory called a {\it{pointer-generator network}} to construct a short summary of a source document by flexibly combining phrases from the source document with words from its existing vocabulary. For each decoder timestep a generation probability $P\mbox{\rm{gen}} \in{} [0, 1]$ is calculated, which weights the probability of {\it{generating}} words from the vocabulary, versus {\it{copying}} words from the source text. The vocabulary distribution and the attention distribution are weighted and summed to obtain the final distribution from which we make our prediction. Note that out-of-vocabulary article words such as "2-0" are included in the final distribution. \emdash{} adapted from See~\etal{}~\cite{SeeetalACL-17}. See Figure~\ref{Context_Episodic_Memory_Dialog} for an application of pointer-generator networks to incorporating episodic memory in dialogue contexts.}
  \label{fig_SeeetalACL-17_Figure_03}
\end{figure}


One approach involves representing a program as an abstract syntax tree and performing a series of repairs that involve replacing complete subtrees in the AST. It might be feasible to use some variant of the pointer-network concept, e.g., {\cite{BhoopchandetalICLR-17}}, {\cite{SeeetalACL-17}} and  {\cite{WangandJiangICLR-17}} or neural programmer framework {\cite{NeelakantanetalICLR-17}}, but there are limitations with all of the alternatives I've run across so far, requiring additional innovation to deal with the dynamic character of editing AST representations, but at least the parsing problem is solved for us \emdash{} all we have to do is make sure that our edits maintain syntactic well-formedness.

Most of the existing pointer-network applications analyze / operate on a fixed structure such as a road map, e.g., examples include the planar graphs that Oriol Vinyals addresses in his paper~\cite{VinyalsetalNIPS-15}, recognizing long-range dependencies in code repositories~\cite{BhoopchandetalICLR-17}, and annotating text to support summarization~\cite{SeeetalACL-17}. Student projects focusing on program-repair might try ingesting programs using an LSTM, creating a pointer-network / DNC-like representation of the AST and then altering selected programs by using fragments from other programs, but be advised this approach will likely require inventing extensions to existing pointer-network techniques.

One possibility for training data is to use the ETH / SRI Python {\urlh{https://www.sri.inf.ethz.ch/py150}{dataset}} that was developed by Veselin Raychev as part of his {\urlh{https://www.sri.inf.ethz.ch/raychev_thesis.pdf}{thesis}} on automated code synthesis. Possible projects include designing a rewrite system for code synthesis based on NLP work from Chris Manning's lab led by Abigail See focusing on text summarization leveraging pointer networks \emdash{} see Figure~\ref{fig_SeeetalACL-17_Figure_03} for a schematic description of the model from their paper~\cite{SeeetalACL-17}. Further afield are program synthesis papers that work starting from specifications like Shin~\etal{}~\cite{ShinetalICLR-18b} out of Dawn Song's lab or recent work from Rishabh Singh and his colleagues~\cite{WangetalCoRR-17}.

Another possibility is to use RL to learn repair rules that operate directly on the AST using various strategies. It's not necessary in this case to represent the AST as a pointer network, but, rather, take the expedient of simply creating a new embedding edited AST after each repair. We can generate synthetic data by taking correct programs from the ETH / SRI dataset and introducing bugs and then use these to generate a reward signal, with harder problems requiring two or three separate repairs. 

It might also be worth exploring the idea of working with program embedding vectors in a manner similar to performing arithmetic operations on word vectors in order to recover analogies \emdash{} see the analysis of Levy and Goldberg~\cite{LevyandGoldbergCONIL-14} in which they demonstrate that analogy recovery is not restricted to simple neural word embeddings. For example, given the AST for a program $P$ with subtree $Q$ and two possible repairs that correspond to replacing $Q$ with either $R$ or $R'$, can we determine which is the better outcome $A = P - Q + R$ or $A' = P - Q + R'$ and might it serve as a distal reward signal to expedite training?

I also recommend Reed and de Freitas~\cite{ReedandDeFreitasCoRR-15} for its application of the idea of using dynamically programmable networks in which the activations of one network become the weights (program) of another network.  The authors note that this approach was mentioned in Sigma-Pi units section of Rumelhart~\etal{}~\cite{RumelhartetalPDP-86b}, appeared in Sutskever and Hinton~\cite{SutskeverandHintonNIPS-09} in the context of learning higher order symbolic relations and in Donnarumma~\etal{}~\cite{DonnarummaetalIJNS-15} as the key ingredient of an architecture for prefrontal cognitive control.


\subsection{Resources}


\begin{itemize}
\item Michael Graziano's {\urlh{https://web.stanford.edu/class/cs379c/calendar_invited_talks/lectures/04/10/index.html}{presentation}} on machines that incorporate an internal model of what consciousness is and attribute that model to themselves and others to make predictions about human behavior~\cite{GrazianoFiRAI-17}.
\item Randall O'Reilly's {\urlh{https://web.stanford.edu/class/cs379c/calendar_invited_talks/lectures/04/12/index.html}{presentation}} on learning mechanisms that rely on a computational model of the prefrontal cortex to control both itself and other brain areas in a strategic, task-appropriate manner~\cite{OReillyandFrankNC-06}.
\item Jay McClelland's {\urlh{https://web.stanford.edu/class/cs379c/calendar_invited_talks/lectures/04/19/index.html}{presentation}} on complementary learning systems that avoid catastrophic forgetting and support the stable learning of new knowledge and learning with imbalanced class labels~\cite{SprechmannetalICLR-18}.
\item Matt Botvinick's {\urlh{https://web.stanford.edu/class/cs379c/calendar_invited_talks/lectures/04/26/index.html}{presentation}} describing a new model of reward-based learning in which a traditional dopamine system trains the prefrontal cortex to operate as its own free-standing learning system~\cite{WangetalNATURE-NEUROSCIENCE-18}.
\end{itemize}


%% file: inputs/03_Interactions.tex

\section{Interaction: Natural Language Processing}


The brain didn't evolve to accommodate language, rather, language evolved to accommodate the brain~\cite{ChaterandChristiansenHLB-11}. Biological and cognitive constraints determine what types of linguistic structure are learned, processed and transmitted from person to person and generation to generation. Language acquisition has comparatively little to do with linguistics and is probably best viewed as a form of skill acquisition. Indeed, we are constantly processing streams of sensory information into successively more abstract representations while simultaneously learning to recode the compressed information into hierarchies of skills that serve our diverse purposes~\cite{ChaterandChristiansenCOiBS-18,ChateretalJML-16}.

Contrary to what some textbook authors might think, students learn to code by writing programs, a process that can be considerably accelerated by timely communication with peers and invested collaborators. In the case of unequal skill levels, communication tends to be on the terms of the more capable interlocutor, and the onus of understanding on the less capable partner in the collaboration. To sustain the collaboration, we need to bootstrap the apprentice to achieve a reasonble threshold level of competence in both language and in working with computers so as to compensate for expert programmer's investment in effort. From a value-proposition perspective, the apprentice has to provide net positive benefit to the programmer from day one.

It is important to keep in mind that any program specification whether in the form of input/output pairs or natural language descriptions when communicated between individuals of differing expertise is just the beginning of a conversation. Experts are often careless in specifying computations and assume too much of the student. Students, on the other hand, can surprise experts with their logical thinking while frustrate and disappoint with their difficulty in handling ambiguity and analogy, particularly of the esoteric sort familiar to professional programmers. Input from an expert will typically consist of a stream of facts, suggestions, heuristics, shortcuts, etc., peppered with clarifications, interruptions and other miscellany.

We propose a hybrid system for achieving competence in communicating programming knowledge and collaboratively generating software by creating a non-differentiable conventional dialogue management system that works in tandem with a differentiable neural-network dialogue system (NDS) that will become more competent as it gains experience coding and collaborating with its expert partner. The deployment of these two language systems will be controlled on a sentence-by-sentence basis by a meta-reinforcement learning (MRL) system that will depend less and less on the more conventional system but likely never entirely eclipse its utility. The MRL system subsumes the role of a beam search or softmax layer in an encoder-decoder dialogue model. 


The conventional system will be built as a hierarchical planner following in the footsteps of the CMU Ravenclaw Dialogue System~\cite{BohusandRudnickyCSL-09} and will come equipped with a relatively sophisticated suite of hierarchical dialogue plans for dealing with communication problems arising due to ambiguity and misunderstanding. While clumsy compared to how humans handle ambiguity and misunderstanding, these dialogue plans are designed to resolve the ambiguity and mitigate the consequences of misunderstanding quickly and get the conversation back on track by attempting various repairs involving requests for definition, clarification, repetition and restatement in as inconspicuous manner as possible~\cite{BohusPhD-07}.

The conventional dialogue system will also include a collection of hierarchical plans for interpreting requests made by the expert programmer to alter programs in the shared editor space, execute programs on specified inputs and perform analyses on the output generated by the program, debugger and other tools provided in the integrated development environment (IDE) accessible through a set of commands implemented as either primitive tasks in the non-differentiable hierarchical planner or through a differentiable neural computer (DNC) interface using the NDS system that will ultimately replace most of the functions of the hierarchical-planner-based dialogue management system.

This dual mode dialogue system and its MRL controller allows the apprentice to practice on its own and rewards it for learning to emulate the less-flexible hierarchical-planner implementation. Indeed there is quite a bit that we can do to preload the apprentice's basic language competence and facility using the instrumented IDE and related compiler chain. A parallel dialogue system implemented using the same hierarchical planner can be designed to carry out the programmer's side of the conversation so as to train the NDS system and the meta-reinforcement learning system that controls its deployment utterance by utterance. We can also train domain-specific language models using $n$-gram corpora gleaned from discussions between pair programmers engaged in writing code for projects requiring the same programming language and working on similar programming tasks.


\subsection{Planning}


In a collaboration, figuring out what to say requires planning and a certain degree of imagination. Suppose you are the apprentice and you want to tell the programmer with whom you're collaborating that you don't understand what a particular expression does. You want to understand what role it plays in the program you are jointly working on. How do you convey this message? What do you need to say explicitly and what can be assumed common knowledge? What does the programmer know and what does she need to be told in order to provide you with assistance?

Somehow you need to model what the programmer knows. In planning what to say, you might turn this around and imagine that you're the programmer and ask how you would respond to an apprentice's effort to solicit help, but in imagining this role reversal you have be careful that you don't assume the programmer knows everything that you do. You need a model of what you know as well as a model of what the programmer knows. This is called Theory of Mind (ToM) reasoning and learning how to carry out such reasoning occurs in a critical stage of child development.

Shared knowledge includes general knowledge about programming, knowledge about the current state of a particular program you are working on, as well as specific details concerning what you are attending to at the moment, including program fragments and variable names that have been mentioned recently in the discussion or can be inferred from context. This sort of reasoning can be applied recursively if, for example, the apprentice wants to know what the programmer thinks it knows about what the apprentice knows. To a large extent we can finesse the problem of reasoning about other minds by practicing transparency, redundancy and simplicity so that both parties can depend on not having to work hard to figure out what the other means. However, there are some opportunities in the programmer's apprentice problem for applying ToM reasoning to parts of the problem that cannot be so easily finessed.

Suppose that the apprentice has started a new program using an existing program $P$ following a suggestion by the expert programmer. Realizing that the body of a loop in $P$ is irrelevant to the task at hand, the apprentice replaces the useless body $B$ with a fragment from another program that does more or less what is required and then makes local changes to the fragment to create a new body $B'$ so that it works with the extant loop variables, e.g., loop counter, termination flag, etc. When the assistant has completed these local changes, the programmer intervenes and changes the name of a variable in $B'$. What induced the programmer to make this change?

The programmer noticed that the variable in $B'$ was not initialized or referenced in $P$ but that another variable that was initialized in $P$ and is no longer referenced \emdash{} it only appeared in the original loop body $B$, is perfectly suited for the purposes of the new program. Assume for the sake of this discussion, that the programmer does not explain her action. How might the assistant learn from this intervention or, at the very least, understand why it was made? A reasonable theory of mind might assume that agents perform actions for reasons and those reasons often have to do with preconditions for acting in the world, and, moreover, that determining if action-enabling preconditions are true often requires effort. A useful ToM also depends on having a model allowing an agent to infer how preconditions enable actions by working backward from actions to enabling preconditions. 


Imagine the following scene, there's a man holding the reins of a donkey harnessed to a two-wheeled cart \emdash{} often called a {\it{dray}} and its owner referred to as a {\it{drayman}} \emdash{} carrying a load of rocks. He makes the donkey rear up and by so doing the surface of the cart tilts, dumping the rocks onto the road which was clearly his intention given the appreciative nods from the onlooking pedestrians. This short {\urlh{https://web.stanford.edu/class/cs379c/resources/amanuensis/content/Donkey_Cart_Draymans_Quick_Unloading_Trick.mp4}{video}} illustrates that, while this might seem an unusual way of delivering a load of rocks, most people think they understand exactly how it was done. Not so!

The fact is that, as with so many other perceptual and conceptual tasks, people feel strongly that they perceived or understood much more than in fact they did. For example, most people would be hard-pressed to induce a donkey to rear up and, if you asked them to draw the donkey harnessed to the cart with its load of stone, they would very likely misrepresent the geometric relationships involving the height of the donkey, how the harness is attached, how far off the ground the axle is located, the diameter of the wheels and the level of the cart surface and center of gravity of the load with respect to the axle's frame of reference. In other words, they would not have \emdash{} and possibly never could have \emdash{} designed a working version of the system used by the drayman.

Now imagine that the drayman has a new apprentice who was watching the entire scene with some concentration, anticipating that he might want to do the very same thing before the first week of his apprenticeship is complete. Sure enough, the next day the drayman tells the apprentice to take a load of bricks to a building site in town where they are constructing a chimney on a new house. He stacks the bricks in a pile that looks something like how he remembers the rocks were arranged on the dray the day before. Unfortunately the load isn't balanced over the axle and almost lifts the donkey off its feet. After some experimentation he discovers how to balance the weight so the donkey can pull the load of bricks without too much effort.

When he finally gets to the building site, he nearly gets trampled by the donkey in the process of repeatedly trying to induce the distressed animal to rear up on its hind legs. Finally, one of the brick masons intervenes and demonstrates the trick. Unfortunately, the bricks don't slide neatly off the dray as the rocks did for the experienced drayman the day before, but instead the bricks on the top of the stack tumble to the pavement and several break into pieces. The helpful brick mason suggests that in the future the assistant should prepare the dray by sprinkling a layer of sand on the surface of cart so that the bricks will slide more freely and that he should also dump the bricks on a softer surface to mitigate possible breakage. He then helps the assistant to unload the rest of the bricks but refuses to pay for the broken ones, telling the assistant he will probably have to pay the drayman to make up for the difference.

An interesting challenge is to develop a model based on what is known about the human brain explaining how memories of the events depicted in the video and extended in the above story might be formed, consolidated, and, subsequently, retrieved, altered, applied and finally assigned a value taking into account the possible negative implications of damaged goods and destroyed property. In the story above, the assistant initially uses his innate "physics engine" to convince himself that he understands the lesson from the master drayman, he then uses a combination of his physical intuitions and trial-and-error to load the cart, but runs up against a wall due to his unfamiliarity with handling reluctant beasts of burden. Finally, he gets into trouble with laws of friction and the quite-reasonable expectations of consumers unwilling to pay for damaged goods. 


We don't propose to solve the general problems of theory-of-mind and physics-based reasoning in developing a programmer's apprentice, though the application provides an interesting opportunity to address particular special cases. As mentioned earlier, the stream of conversation between the assistant an expert programmer will inevitably relate to many different topics and specialized areas of expertise. It will include specific and general advice, reasons for acting, suggestions for what to attend to and a wide range of comments and criticisms. Several recent approaches for combining planning and prediction, especially in the case of partially observable Markov decision processes, are particularly promising for this application~\cite{GroshevetalCoRR-17,FinnetalCoRR-17,FoersteretalCoRR-17,WayneetalCoRR-18,SilveretalICML-17}.

The apprentice will want to separate this information into different categories to construct solutions to problems that arise at multiple levels of abstraction and complexity during code synthesis. Or will it? We like to think of knowledge neatly packaged into modules that result in textbooks, courses, monographs, tutorials, etc. The apparent order in which activities appear in a stream of activities is largely a consequence of the context in which those activities are carried out. They may seem to arise in accord with some plan, as if assembled and orchestrated with a particular purpose in mind, but, even if there was plan at the outset, we tend to make up things on the fly to accommodate the sort of unpredictable circumstances that characterize most of our evolutionary history. 

In some cases that context or purpose is used to assign a name, but that name or contextual handle is seldom used to initiate or identify the activity except in academic circumstances where divisions and boundaries are highly prized and made much of. The point of this is that in a diverse stream of activities \emdash{} or utterances intended to instigate activities \emdash{} credit assignment can be difficult. Proximity in terms of the length of time or number of intervening activities between a action and a reward is not necessarily a good measure of its value. We suggest it is possible to build a programer's apprentice or other sort of digital assistant that performs its essential services primarily by learning to predict actions, their consequences and their value from observing such a diverse stream of dialog intermixed with actions and observations. 


\subsection{Hybrids}

When the programmer tells the assistant to replace the name of a variable in one location in a program with the name of a variable in another location, the process starts with a contextually rich representation in the programmer's brain corresponding to the activation of millions or billions of neurons in circuits distributed broadly throughout the cortex. This pattern of activation is compressed into a more compact representation used to generate a sequence of words uttered one at a time as if condensing out of a cloud of commingled thoughts in droplets or phrasal showers uttered in sudden bursts of words that are \emdash{} ignoring the intervening stages of speech production in the programmer and auditory processing in the assistant \emdash{} subsequently converted into activations in peripheral subnetworks of the assistant and quickly propagate to other subnetworks throughout the assistant's neural-network architecture. 

The resulting activations insinuate fractal patterns of meaning into broadly distributed subnetworks of the apprentice subtly altering activity in some and substantially altering activity in others, contributing to the formation of another contextually rich representation in the apprentice's brain. The imperative conveyed by the programmer's tone of voice produces a quick response. The apprentice performs a sequence of well rehearsed steps that involve activating a sequence of patterns in the non-differentiable interface connecting the assistant to the integrated development environment. This sequence is produced by recurrent networks operating much like the programmer's speech production circuits. The resulting patterns produce a sequence of unambiguous words \emdash{} requiring no additional context to interpret, that immediately produce the desired change and are displayed on the screen shared by the programmer and the apprentice.

What does this combination of expert-human biological computing, differentiable connectionist models and non-differentiable symbolic systems buy us? The human expert provides heuristic advice and technical guidance. The connectionist components enable natural language communication between human and machine and enable the system to discover and exploit structure in computer programs to facilitate code synthesis and reduce brute search. Finally, the symbolic components allow the connectionist components \emdash{} and, by extension, the human expert \emdash{} to directly engage with computers as prosthetic extensions in which compiling and running code is as natural as playing video games.


\subsection{Resources}

\begin{itemize}
\item Neil Rabinowitz's {\urlh{https://web.stanford.edu/class/cs379c/calendar_invited_talks/lectures/04/17/index.html}{presentation}} on learning a machine theory-of-mind model that relies on meta-learning to build mental models of the agents that it encounters from observations of their behaviour alone~\cite{RabinowitzetalCoRR-18}.
\item Greg Wayne's {\urlh{https://web.stanford.edu/class/cs379c/calendar_invited_talks/lectures/05/03/index.html}{presentation}} on {\tt{MERLIN}} a method for prediction in environments corresponding to partially observable Markov decision processes in which memory formation is guided by predictive modeling~\cite{WayneetalCoRR-18}.
\item Oriol Vinyals' {\urlh{https://web.stanford.edu/class/cs379c/calendar_invited_talks/lectures/05/10/index.html}{presentation}} on an approach for model-based plan construction, evaluation and execution applied to sequential decision making problems relying on a method of imagination-based forecasting~\cite{PascanuetalCoRR-17}.
\item Devi Parikh's {\urlh{https://web.stanford.edu/class/cs379c/calendar_invited_talks/lectures/05/22/index.html}{public lectures}} on learning to conduct meaningful dialog with humans in natural, conversational language by grounding the conversation in shared visual experience, inferring its context from history~\cite{DasetalCVPR-17}.
\end{itemize}


%% file: inputs/04_Production.tex

\section{Generation: Automated Code Synthesis}


Gulwani~\etal{}~\cite{GulwanietalFaTiPL-17} provide an up-to-date survey of challenges and technologies in automated program synthesis. Machine learning is allocated only 10 of the more than 100 pages in the review with neural network-methods singled out as a separate section of 6 additional pages. My experience is that many current devotees are largely ignorant concerning the relevant history including the many successful applications for handling interesting special cases. I won't attempt to remedy that state of affairs here except to recommend that readers interested in program synthesis spend the time to familiarize themselves with the relevant work including both successes and notable failures. 

For those of you primarily interested in neural-network methods, understanding this background is likely to prove useful in developing hybrid systems that combine connectionist models and more conventional symbolic methods. Already, the power of high-dimensional vector-space representations, context-sensitive embedding spaces and fully-differentiable models trained with backpropagation is winning converts among advocates of more traditional methods, even as the latest generation of neural-network experts working on what is popularly called neural program synthesis are rediscovering some of the same special cases that have been exploited in deductive and constraint-based approaches to automated code synthesis.


\subsection{Teaching}


Given the title, you might have expected that this document would be all about automatic programming, but it's really about human augmentation and human-computer collaboration. Programming is all about representing procedural knowledge in an interpretable form, i.e., executable on some computational substrate. Teaching someone to program or, for that matter, teaching someone to do most anything nontrivial, is also about representing and communicating procedural knowledge in an interpretable form \emdash{} the clearer and more precise the communication, the simpler and less knowledgeable the required substrate.

In this section, we attempt to identify special cases in which it is relatively easy for a human programmer to instruct an AI system how to facilitate the conversion of thoughts conveyed in natural language into working programs \emdash{} see Graham Neubig's CS379C calendar {\urlh{https://web.stanford.edu/class/cs379c/calendar_invited_talks/lectures/05/01/index.html}{page}} for a sample of his NLP work on code generation and learning dialog systems. We are not anticipating that we will be able to read minds so much as exploit shared knowledge in order to translate succinct descriptions of desired computations \emdash{} along with an implicit understanding of what constitutes a suitable computational solution \emdash{} into working software possibly with guarantees of its correctness and performance.

How does this perspective inform the discussion in this section? Prior to any training the apprentice comes equipped with a very basic language facility and an innate ability to work with computer programs, but the former is practically useless since the meta-learning controller hasn't been trained and the latter only produces results in the same way that a baby has virtually no control over its limbs and torso. It learns to communicate by reinforcement when it successfully carries out a command from the expert programmer. A built-in training system expedites this process and relieves some of the burden from the programmer by generating synthetic commands that serve to initialize the meta-learning controller allowing the assistant to achieve a basic facility for directly translating natural language commands exercising the integrated development environment.

The apprentice also comes pre-trained with a (semantic) {\urlh{https://en.wikipedia.org/wiki/Language_model}{language model}} trained on a corpus that includes a large vocabulary of practical and technical terms used by programmers in talking about programs and programming. The apprentice ingests a large corpus of programs and program fragments written in the language that it will use for writing new programs, resulting in an embedding space that encodes these programs and an encoder-decoder translation facility that allows it to read and write programs represented as abstract syntax trees. Were there more computer- / natural-language {\urlh{https://en.wikipedia.org/wiki/Parallel_text\#Parallel_corpora}{parallel corpora}}, supervised training of systems that depend on natural-language specification as input would be tempting. As it is, we have to rely on more subtle strategies.


Perhaps it is not surprising that many current neural-network approaches take advantage of neural-network technology originally designed for machine translation, question answering and related natural-language processing applications. Programs are linguistic objects with relatively simple syntax. Their syntax is well understood since linguistically adept computer scientists designed them. Parsing programs is trivial using standard compiler tools. Their semantics is revealed by running them, and not just their input-output behavior \emdash{} execution traces can be used to reveal the meaning of every function and fragment. 

Programming languages are syntactically unforgiving \emdash{} a source of aggravation for beginning programmers, but we have sophisticated editors and syntax checkers that avoid most problems and there is no reason not to build them into the integrated development environment used by the programmer's assistant and shared with the programmer. Internally, the assistant can work with equivalent abstract syntax trees making it easy to ingest, embed, manipulate and generate proposals for performing program transformations on such representations. Human readable code can be recovered for the programmer's convenience.

Successful code synthesis can be verified by running the program and comparing with the provided input-output examples. Modulo the intractability of the halting problem, failure for relatively simple programs can be easily determined. Program {\urlh{https://en.wikipedia.org/wiki/Correctness_(computer_science)}{correctness}} is generally specified with respect to a specification and hence is more difficult to pin down, though the term generally implies the existence of a mathematical proof and, hence, one formal-methods approach to code synthesis involves generating a {\urlh{https://en.wikipedia.org/wiki/Constructive_proof}{constructive proof}} of correctness. 

Semantic embeddings are increasingly common~\cite{DevlinetalICLR-18,ChistyakovetalICLR-17,WangetalCoRR-17,XuetalCoRR-17,PiechetalICML-15} based on {\urlh{https://en.wikipedia.org/wiki/Software_diagnosis\#Characteristics}{execution traces}}, {\urlh{https://en.wikipedia.org/wiki/Log_file\#Event_logs}{event logs}}, or {\urlh{https://en.wikipedia.org/wiki/Invariant_(computer_science)}{program invariants}}. See this discussion log {\urlh{https://web.stanford.edu/class/cs379c/class_messages_listing/index.html\#program_embedding_core_technology}{entry}} and Rishabh Singh's CS379C calendar {\urlh{https://web.stanford.edu/class/cs379c/calendar_invited_talks/lectures/05/24/index.html}{page}} plus this discussion log {\urlh{https://web.stanford.edu/class/cs379c/class_messages_listing/index.html\#execution_logs_program_traces}{entry}} and Dawn Song's CS379C calendar {\urlh{https://web.stanford.edu/class/cs379c/calendar_invited_talks/lectures/05/31/index.html}{page}} for more on semantic embeddings. I'm not going into detail here for the simple reason that it is still early days and your best approach to learn more is to read the papers, check out the slides and watch the videos mentioned in this paragraph and at the end of this section \emdash{} you might also take a look at Danny Tarlow's list of selected program synthesis papers {\urlh{./content/Selected_Program_Synthesis_Papers_Tarlow.pdf}{here}}.


\subsection{Projects}


In the Spring 2018 instance of Stanford {\urlh{https://web.stanford.edu/class/cs379c/}{CS379C}}, 30 students, organized in 12 teams proposed and carried out projects relating to the programmer's apprentice. As an example, the team consisting of Marcus Gomez, Nate Gruver, Michelle Lam, Rohun Saxena and Lucy Wang~\cite{CS379C_Final_Project_Gomezetal-18} decided based on their coding habits that the most helpful assistant would be one that could be trained to understand design through the structure of an application programming interface ({\urlh{https://en.wikipedia.org/wiki/Application_programming_interface}{API}}) and assist in the completion of the code through a divide-and-conquer approach {\urlh{./content/CS379C_Final_Project_Gomezetal-18.pdf}{PDF}}.


\begin{figure}
  \begin{center}
    \includegraphics[width=5.0in]{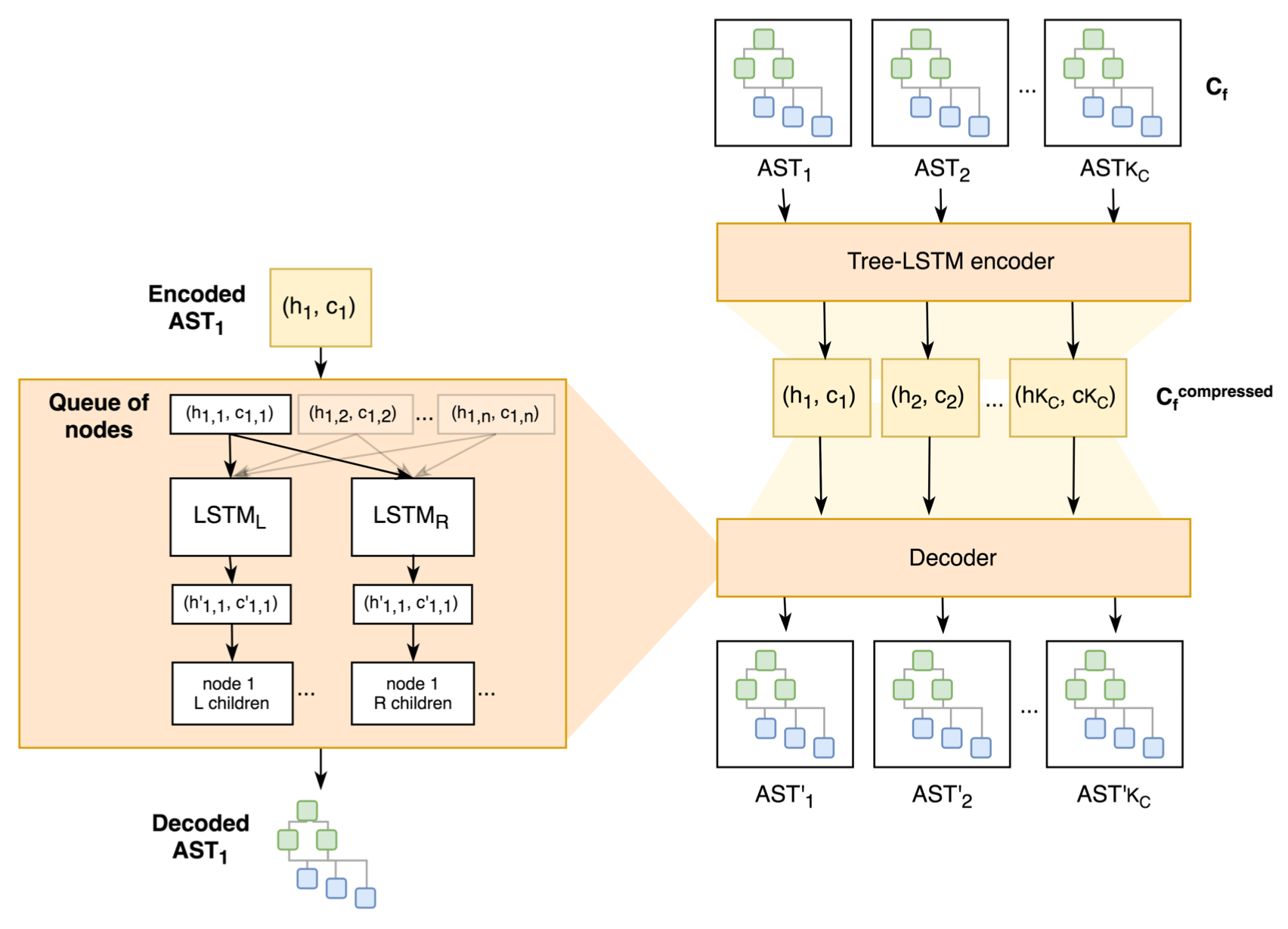}
  \end{center}
  \caption{Summary of the proposed LSTM-based state-embedding architecture for $C_{f}$ described in Gomez~\etal{}~\cite{CS379C_Final_Project_Gomezetal-18}. The code representation, $C_{f}$, is by default variable in size since the abstract syntax trees that comprise the individual helper functions are variable size. The authors solve this problem by training an autoencoder penalized on reconstruction loss such that the decoded output is the binary tree representation of the original AST.}
  \label{fig_CS379C_Final_Project_Gomezetal_Figure_03}
\end{figure}


The system they envision would be designed to integrate seamlessly into programmer's workflow, minimize interaction and start from a simple header file and dependency graph. They took pains to standardize the format of the input in order to simplify learning a compressed fixed-size state vector as an ordered list of GLoVe vectors~\cite{PenningtonetalEMNLP-14} \emdash{} see Figure~\ref{fig_CS379C_Final_Project_Gomezetal_Figure_03}. They employ the method of {\it{generative adversarial imitation learning}} (GAIL) developed by Ho and Ermon~\cite{HoandErmonCoRR-16} to extract a policy from data as if it were obtained by reinforcement learning following inverse reinforcement learning. GAIL should enable them to derive a model-free imitation-learning algorithm that obtains significant performance gains over existing model-free methods in imitating complex behaviors in large, high-dimensional environments.


Episodic memory is an important component of the programmer's apprentice. In order to exploit what you've learned through experience and previously consolidated in episodic memory, it is often necessary to reconstruct memories in enough detail that you can compare the past with current experience, determine if activities applied to resolve problems in the past apply in the present, and, if necessary, adapt those early responses to fit the present circumstances. Catherine Wong, in her final project report~\cite{CS379C_Final_Project_Wong-18} for CS379C, developed a new neural-network model for content-based and selectively-reconstructive memory inspired by research on {\it{hippocampal indexing theory}}~\cite{TeylerandRudyHIPPOCAMPUS-07} and {\it{adaptive deconvolutional networks}}~\cite{ZeileretalICCV-11,ZeileretalCVPR-10} {\urlh{./content/CS379C_Final_Project_Wong-18.pdf}{PDF}}. 

In designing her model she built upon the Kavukcuoglu~\etal{}~\cite{KavukcuogluetalNIPS-10} work on {\urlh{http://koray.kavukcuoglu.org/research.html}{convolutional predictive sparse decomposition}} and Kingma and Welling~\cite{KingmaandWellingCoRR-13} on {\urlh{https://www.ibm.com/blogs/research/2018/05/disentanglement-deep-learning/}{explicit feature disentanglement using variational autoencoders}}. While previous work has focused on deciding {\it{how}} to remember~\cite{WayneetalCoRR-18}, Catherine's model emphasizes deciding {\it{what}} to remember, focusing on the problem of leveraging compressive coding to balance the demands of efficient and effective content-based lookup with domain specificity. Her work is also closely related to and consistent with recent research~\cite{LatchoumaneetalNEURON-17,FanetalJCN-17,NielsenNLM-15} on the role of high-frequency {\urlh{https://en.wikipedia.org/wiki/Sleep_spindle}{thalamic sleep spindles}} during sleep-dependent memory consolidation.


Technology for automatic software repair is becoming an important component of software maintenance and automatic code synthesis~\cite{MonperrusACM-17}. Maurice Chiang, Yousef Hindy, Peter Lu, Sophia Sanchez and Michael Smith~\cite{CS379C_Final_Project_Luetal-18} developed a neural network architecture to support error correction in abstract syntax trees that can be used as a code repair module in a more general neural code synthesis system {\urlh{./content/CS379C_Final_Project_Luetal-18.pdf}{PDF}}. Their approach is novel in its use of multiple strategies realized as separate submodules, coordinated by a master controller, so that the whole system is trained end-to-end after training each of the submodules independently. 

The Lu~\etal{} work~\cite{CS379C_Final_Project_Luetal-18} takes advantage of a number of recent innovations in neural code synthesis. Specifically they leverage the Cai~\etal{}~\cite{CaietalICLR-17} work on using recursion to generalize programs to handle novel inputs. They employ a version of Pascanu~\etal{}~\cite{PascanuetalCoRR-17} model-based planning (IBP) that accepts partially complete or incorrect ASTs, extending the IBP controller to coordinate the separate submodules. They also propose a strategy for handling intermediate reward signals to compensate for the relative sparsity of ground truth data and a clever approach to comparing programs that relies on normalizing and comparing stack traces computed from input-output pairs~\cite{BrodieetalICAC-05,GuptaetalPATENT-06,SmithandWatermanJMB-81}. 

These three projects illustrate how ideas from such seemingly disjoint disciplines as cognitive neuroscience and theoretical computer science can come together to create sophisticated new technologies. Unfortunately, competition between disciplines has resulted in convenient lapses in memory exacerbated by strategic internal rebranding of ideas. The current period of increased collaboration illustrates the advantages of creating and maintaining cross-disciplinary ties. This generation of students is being exposed to ideas from a broad range of ideas relating to synthetic and biological computing and will hopefully pass on their interests to their students and more narrowly focused colleagues.


These innovative student projects target important problems that we face in designing complex collaborative AI systems such as the programmer's apprentice. Their proposed solutions highlight the variety and scalabilty of the architectural components now available that allow for the creative manipulation of the ungainly objects that comprise the inputs and outputs of software engineering practice, e.g., natural language specifications of arbitrary format, the encoding, analysis and procedural extraction of programming knowledge from dialog, and dealing with computer programs of arbitrary size and complexity and computed artifacts such as execution traces that defy explicit encoding due to their size and format variability.


The Programmer's Apprentice is {\urlh{https://en.wikipedia.org/wiki/AI-complete}{AI complete}} in the sense that it is equivalent to that of making computers at least as intelligent as human beings. It isn't necessarily the part relating to code synthesis \emdash{} though, depending on how you define fully automatic code synthesis starting from a natural language description, that too could be said to be AI complete. Assuming we are satisfied to build a system capable of {\it{assisting}} rather than {\it{replacing}} a programmer, then the hard problem is in the ability to interact easily with humans.

This is not to say that we can't build highly capable assistants in the near term. We simply have to constrain the problem appropriately, and I believe we can do much better than the current breed of personal assistants in designing a programmer's assistant that can considerably increase the productivity of professional software engineers and enable reasonably competent programmers to become much more effective. The speakers contributing to class discussions and students working on related projects demonstrate that many of the crucial features required of a programmer's assistant are within our grasp.

Such features include speaker- and task-dependent management of episodic memory including memory consolidation, procedural abstraction and control-policy / strategy integration, meta-level control of reinforcement learning and policy application integrating multiple sources of knowledge pertaining to programming and maintaining continuous, problem-focused dialog, and basic programming capabilities including program repair and simplification, identifying and repurposing existing program fragments to generate new programs, and translating natural language program specifications into working programs.

If you are serious about wanting to build some variant of digital amanuensis to assist in writing programs, the good news is that there are a lot of tools you can leverage. There is also some excitement and optimism derived from the scattered successes in applying modern connectionist models to the problem of automatic code synthesis. For those with the perspective to see beyond the new advances, there is the distinct possibility of combining the deep learning methods that have recently shown such promise with the considerable body of work on deductive and statistical methods applied to automatic program synthesis~\cite{GulwanietalFaTiPL-17}. 


\subsection{Resources}

\begin{itemize}
\item Daniel Abolafia's {\urlh{https://web.stanford.edu/class/cs379c/calendar_invited_talks/lectures/04/24/index.html}{presentation}} on iterative optimization for program synthesis in the presence of a reward function over the output of programs, where the goal is to find programs with maximal rewards~\cite{AbolafiaetalCoRR-18}.
\item Graham Neubig's {\urlh{https://web.stanford.edu/class/cs379c/calendar_invited_talks/lectures/05/01/index.html}{presentation}} on a novel neural architecture for parsing natural language descriptions into source code powered by a grammar model to explicitly capture the target syntax as prior knowledge~\cite{YinandNeubigACL-17}.
\item Rishabh Singh's {\urlh{https://web.stanford.edu/class/cs379c/calendar_invited_talks/lectures/05/24/index.html}{presentation}} on using a strong statistical model for semantic code repair to predict bug locations and exact fixes without access to information about the intended correct behavior of the program.~\cite{DevlinetalICLR-18}.
\item Dawn Song and Xinyun Chen's {\urlh{https://web.stanford.edu/class/cs379c/calendar_invited_talks/lectures/05/31/index.html}{presentation}} on program synthesis from input-output examples, tree-to-tree neural networks for program translation, and attention for program synthesis from natural language descriptions~\cite{ChenetalICLR-18b}.
\end{itemize}


%% file: inputs/A_Architecture.tex

\section{Human-Like Cognitive Architectures}


Our objective in developing systems that incorporate characteristics of human intelligence is three fold: First, humans provide a compelling solution to the problem of building intelligent systems that we can use as a basic blueprint and then improve upon. Second, the resulting AI systems are likely to be well suited to developing assistants that complement and extend human intelligence while operating in a manner comprehensible to human understanding. Finally, cognitive and systems neuroscience provide clues to engineers interested in exploiting what we know concerning how humans think about and solve problems. In this appendix, we demonstrate one attempt to concretely realize what we've learned from these disciplines in an architecture constructed from off-the-shelf neural networks. 

\begin{figure}
  \begin{center} 
    \includegraphics[width=275pt]{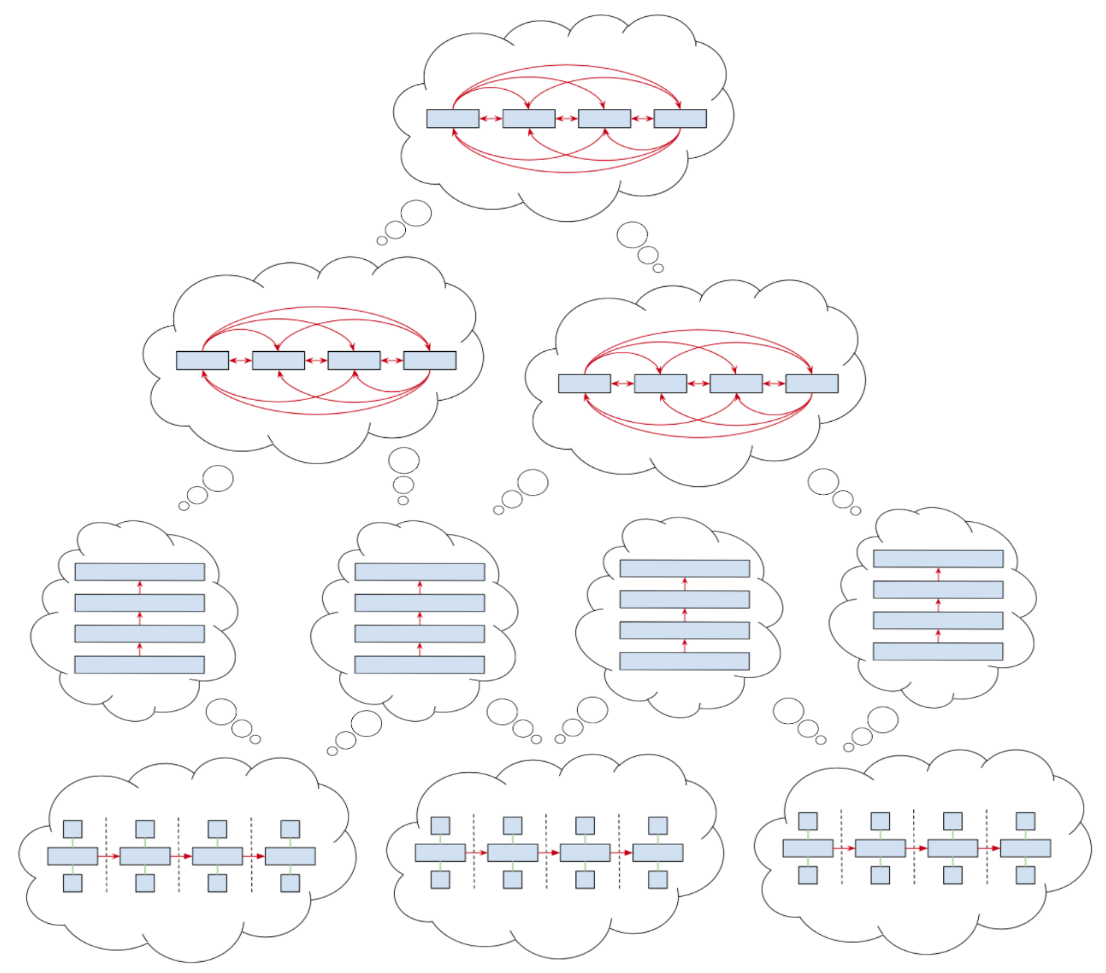} 
  \end{center}
  \caption{The architecture of the apprentice sensory cortex including the layers corresponding to abstract, multi-modal representations handled by the association areas can be realized as a multi-layer hierarchical neural network model consisting of standard neural network components whose local architecture is primarily determined by the sensory modality involved. This graphic depicts these components as encapsulated in thought bubbles of the sort often employed in cartoons to indicate what some cartoon character is thinking. Analogously, the technical term "thought vector" is used to refer to the activation state of the output layer of such a component.}
  \label{fig_Posterior_Cortex_Semantic_Memory}
\end{figure}

The programmer's apprentice relies on multiple sources of input, including dialogue in the form of text utterances, visual information from an editor buffer shared by the programmer and apprentice and information from a specially instrumented integrated development environment designed for analyzing, writing and debugging code adapted to interface seamlessly with the apprentice. This input is processed by a collection of neural networks modeled after the primary sensory areas in the primate brain. The outputs of these networks feed into a hierarchy of additional networks corresponding to uni-modal secondary and multi-modal association areas that produce increasingly abstract representations as one ascends the hierarchy \emdash{} see Figure~\ref{fig_Posterior_Cortex_Semantic_Memory}.
  

Stanislas Dehaene and his colleagues at the Coll\`{e}ge de France in Paris developed a computational model of consciousness that provides a practical framework for thinking about consciousness that is sufficiently detailed for much of what an engineer might care about in designing digital assistants~\cite{Dehaene2014}. Dehaene's work extends the {\it{Global Workspace}} Theory of Bernard Baars~\cite{Baars1988}. Dehaene's version of the theory combined with Yoshua Bengio's concept of a {\it{consciousness prior}} and deep reinforcement learning~\cite{MnihetalCoRR-13,NairetalCoRR-15} suggest a model for constructing and maintaining the cognitive states that arise and persist during complex problem solving~\cite{BengioCoRR-17}.

Global Workspace Theory accounts for both conscious and unconscious thought with the primary distinction for our purpose being that the former has been selected for attention and the latter has not been so selected. Sensory data arrives at the periphery of the organism. The data is initially processed in the primary sensory areas located in posterior cortex, propagates forward and is further processed in increasingly-abstract multi-modal association areas. Even as information flows forward toward the front of the brain, the results of abstract computations performed in the association areas are fed back toward the primary sensory cortex. This basic pattern of activity is common in all mammals. 

\begin{figure}
  \begin{center} 
    \includegraphics[width=325pt]{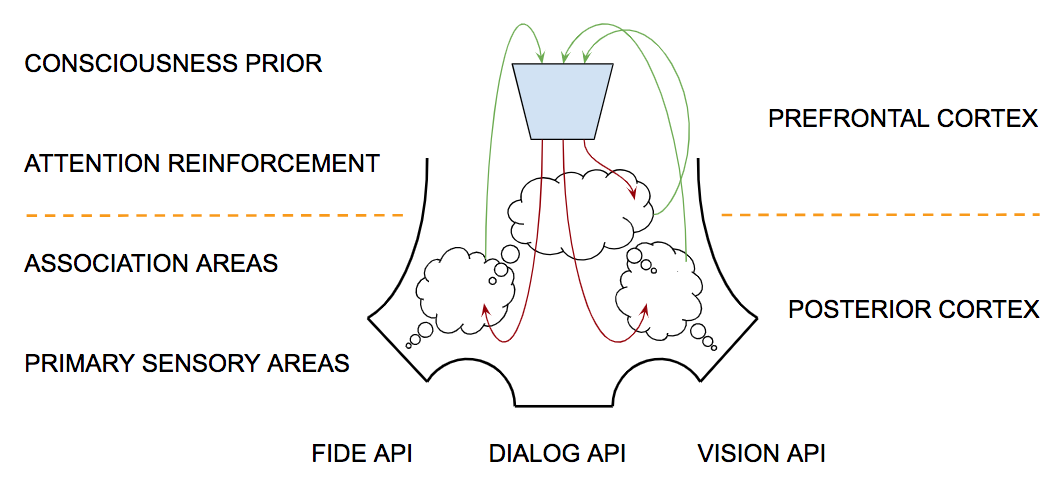} 
  \end{center}
  \caption{The basic capabilities required to support conscious awareness can be realized in a relatively simple computational architecture that represents the apprentice's global workspace and incorporates a model of attention that surveys activity throughout somatosensory and motor cortex, identifies the activity relevant to the current focus of attention and then maintains this state of activity so that it can readily be utilized in problem solving.  In the case of the apprentice, new information is ingested into the model at the system interface, including dialog in the form of text, visual information in the form of editor screen images, and a collection of programming-related signals originating from a suite of software development tools. 
Single-modality sensory information feeds into multi-modal association areas to create rich abstract representations. Attentional networks in the prefrontal cortex take as input activations occurring throughout the posterior cortex. These networks are trained by reinforcement learning to identify areas of activity worth attending to and the learned policy selects a set of these areas to attend to and sustain. This attentional process is guided by a prior that prefers low-dimensional thought vectors corresponding to statements about the world that are either true, highly probable or very useful for making decisions. Humans can sustain only a few such activations at a time. The apprentice need not be so constrained.}
  \label{fig_Global_Workspace_Conscious_Attention}
\end{figure}

The human brain has evolved to handle language. In particular, humans have a large frontal cortex that includes machinery responsible for conscious awareness and that depends on an extensive network of specialized neurons called spindle cells that span a large portion of the posterior cortex allowing circuits in the frontal cortex to sense relevant activity throughout this area and then manage this activity by creating and maintaining the persistent state vectors that are necessary when inventing extended narratives or working on complex problems that require juggling many component concepts at once. Figure~\ref{fig_Global_Workspace_Conscious_Attention} suggests a neural architecture combining the idea of a global workspace with that of an attentional system for identifying relevant input.


Fundamental to our understanding of human cognition is the essential tradeoff between fast, highly-parallel, context-sensitive, distributed connectionist-style computations and slow, serial, systematic, combinatorial symbolic computations. In developing the programmer's apprentice, symbolic computations of the sort common in conventional computing are realized using extensions that provide a differentiable interface to conventional memory and information processing hardware and software. Such interfaces include the Neural Turing Machine~\cite{GravesetalCoRR-14} (NTM), Memory Network Model~\cite{WestonetalCoRR-14,SukhbaataretalCoRR-15} and Differentiable Neural Computer~\cite{GravesetalNATURE-16} (DNC).

The global workspace summarizes recent experience in terms of sensory input, its integration, abstraction and inferred relevance to the context in which the underlying information was acquired. To exploit the knowledge encapsulated in such experience, the apprentice must identify and make available relevant experience. The apprentice's experiential knowledge is encoded as tuples in a Neural Turing Machine (NTM) memory that supports associative recall. We'll ignore the details of the encoding process to focus on how episodic memory is organized, searched and applied to solving problems.

\begin{figure}
  \begin{center} 
    \includegraphics[width=400pt]{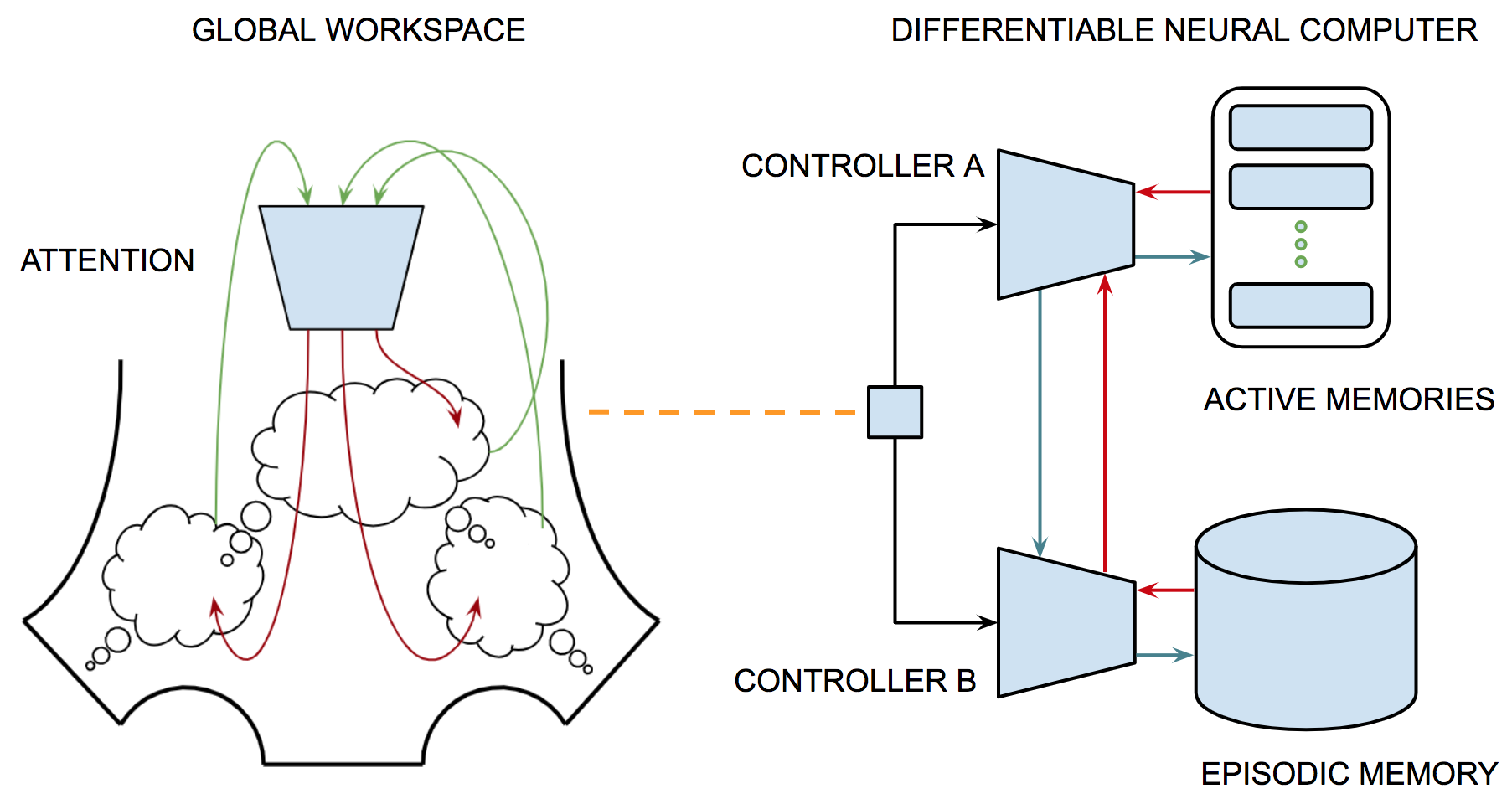} 
  \end{center}
  \caption{You can think of the episodic memory encoded in the hippocampus and entorhinal cortex as RAM and the actively maintained memories in the prefrontal cortex as the contents of registers in a conventional von Neumann architecture. Since the activated memories have different temporal characteristics and functional relationships with the contents of the global workspace, we implement them as two separate NTM memory systems each with its own special-purpose controller. Actively maintained information highlighted in the global workspace is used to generate keys for retrieving relevant memories that augment the highlighted activations. In the DNC paper~\cite{GravesetalNATURE-16} appearing in {\it{Nature}}, the authors point out that "an associative key that only partially matches the content of a memory location can still be used to attend strongly to that location [allowing] allowing the content of one address [to] effectively encode references to other addresses". The contents of memory consist of thought vectors that can be composed with other thought vectors to shape the global context for interpretation.}
  \label{fig_Global_Workspace_Episodic_Memory}
\end{figure}

In the biological analog of an NTM the hippocampus and entorhinal region of the frontal cortex play the role of episodic memory and several subcortical circuits including the basal ganglia comprise the controller~\cite{OReillyetalLEABRA-16,OReillySCIENCE-06}. The controller employs associative keys in the form of low-dimensional vectors generated from activations highlighted in the global workspace to access related memories that are then actively maintained in the prefrontal cortex and serve to bias processing throughout the brain but particularly in those circuits highlighted in the global workspace. Figure~\ref{fig_Global_Workspace_Episodic_Memory} provides a sketch of how this is accomplished in the apprentice architecture. 


\begin{figure}
  \begin{center} 
    \includegraphics[width=375pt]{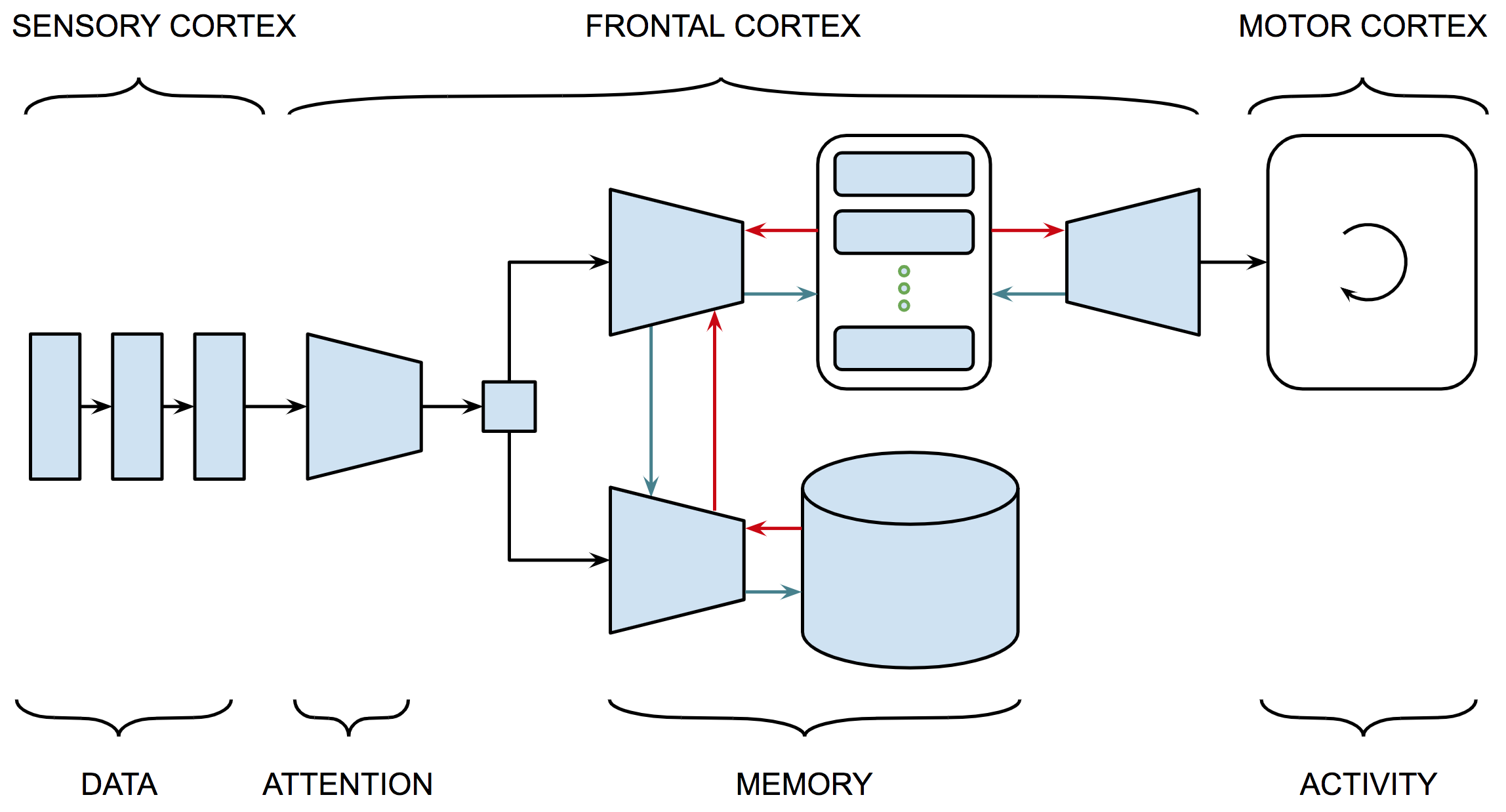} 
  \end{center}
  \caption{This slide summarizes the architectural components introduced so far in a single model. Data in the form of text transcriptions of ongoing dialogue, source code and related documentation and output from the integrated development environment are the primary input to the system and are handled by relatively standard neural network models. The Q-network for the attentional RL system is realized as a multi-layer convolutional network. The two DNC controllers are straightforward variations on existing network models with a second controller responsible for maintaining a priority queue encodings of relevant past experience retrieved from episodic memory. The nondescript box labeled "motor cortex" serves as a placeholder for the neural networks responsible for managing dialogue and handling tasks related to programming and code synthesis.}
  \label{fig_High_Level_Assistant_Architecture}
\end{figure}

Figure~\ref{fig_High_Level_Assistant_Architecture} combines the components that we've introduced so far in a single neural network architecture. The empty box on the far right includes both the language processing and dialogue management systems as well the networks that interface with FIDE and the other components involved in code synthesis. There are several classes of programming tasks that we might tackle in order to show off the apprentice, including commenting, extending, refactoring and repairing programs. We could focus on functional languages like Scheme or Haskell, strongly typed languages like Pascal and Java or domain specific languages like HTML or SQL. 

However, rather than emphasize any particular programming language or task, in the remainder of this appendix we focus on how one might represent structured programs consisting of one or more procedures in a distributed connectionist framework so as to exploit the advantages of this computational paradigm. We believe the highly-parallel, contextual, connectionist computations that dominate in human information processing will complement the primarily-serial, combinatorial, symbolic computations that characterize conventional information processing and will have a considerable positive impact on the development of practical automatic programming methods.


The integrated development environment and its associated software engineering tools constitute an extension of the apprentice's capabilities in much the same way that a piano or violin extends a musician or a prosthetic limb extends someone who has lost an arm or leg. The extension becomes an integral part of the person possessing it and over time their brain creates a topographic map that facilitates interacting with the extension. This was early recognized in the work of Hubel and Wiesel~\cite{HubelandWieselJoP-68,HubelandWieselJoP-62} on the striate cortex of the cat and macaque monkey and in the work of Wilder Penfield~\cite{PenfieldandBoldreyBRAIN-37} on the primary motor and somatosensory areas in the human brain.
 
As engineers designing the apprentice, part of our job is to create tools that enable the apprentice to learn its trade and eventually become an expert. Conventional IDE tools simplify the job of software engineers in designing software. The fully instrumented IDE (FIDE) that we engineer for the apprentice will be integrated into the apprentice's cognitive architecture so that tasks like stepping a debugger or setting breakpoints are as easy for the apprentice as balancing parentheses and checking for spelling errors in a text editor is for us.

As a first step in simplifying the use of FIDE for coding, the apprentice is designed to manipulate programs as abstract syntax trees (AST) and easily move back and forth between the AST representation and the original source code in collaborating with the programmer. Both the apprentice and the programmer can modify or make references to text appearing in the FIDE window by pointing to items or highlighting regions of the source code. The text and AST versions of the programs represented in the FIDE are automatically synchronized so that the program under development is forced to adhere to certain syntactic invariants. 


\begin{figure}
  \begin{center} 
    \includegraphics[width=225pt]{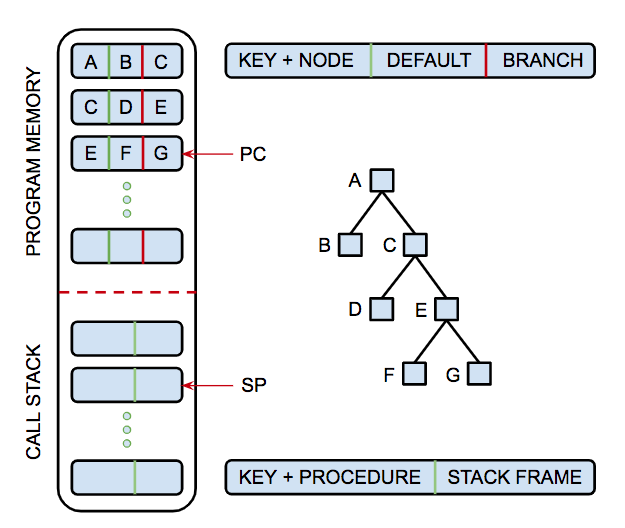} 
  \end{center}
  \caption{We use pointers to represent programs as abstract syntax trees and partition the NTM memory, as in a conventional computer, into program memory and a LIFO execution (call) stack to support recursion and reentrant procedure invocations, including call frames for return addresses, local variable values and related parameters. The NTM controller manages the program counter and LIFO call stack to simulate the execution of programs stored in program memory. Program statements are represented as embedding vectors and the system learns to evaluate these representations in order to generate intermediate results that are also embeddings. It is a simple matter to execute the corresponding code in the FIDE and incorporate any of the results as features in embeddings.}
  \label{fig_Differentiable_Structured_Programs}
\end{figure}


To support this hypothesis, we are developing distributed representations for programs that enable the apprentice to efficiently search for solutions to programming problems by allowing the apprentice to easily move back and forth between the two paradigms, exploiting both conventional approaches to program synthesis and recent work on machine learning and inference in artificial neural networks. Neural Turing Machines coupled with reinforcement learning are capable of learning simple programs. We are interested in representing structured programs expressed in modern programming languages. Our approach is to alter the NTM controller and impose additional structure on the NTM memory designed to support procedural abstraction. 

What could we do with such a representation? It is important to understand why we don't work with some intermediate representation like bytecodes. By working in the target programming language, we can take advantage of both the abstractions afforded by the language and the expert knowledge of the programmer about how to exploit those abstractions. The apprentice is bootstrapped with several statistical language models: one trained on a natural language corpus and the other on a large code repository. Using these resources and the means of representing and manipulating program embeddings, we intend to train the apprentice to predict the next expression in a partially constructed program by using a variant of imagination-based planning~\cite{PascanuetalCoRR-17}. As another example, we will attempt to leverage NLP methods to generate proposals for substituting one program fragment for another as the basis for code completion. 


\begin{figure}
  \begin{center} 
    \includegraphics[width=200pt]{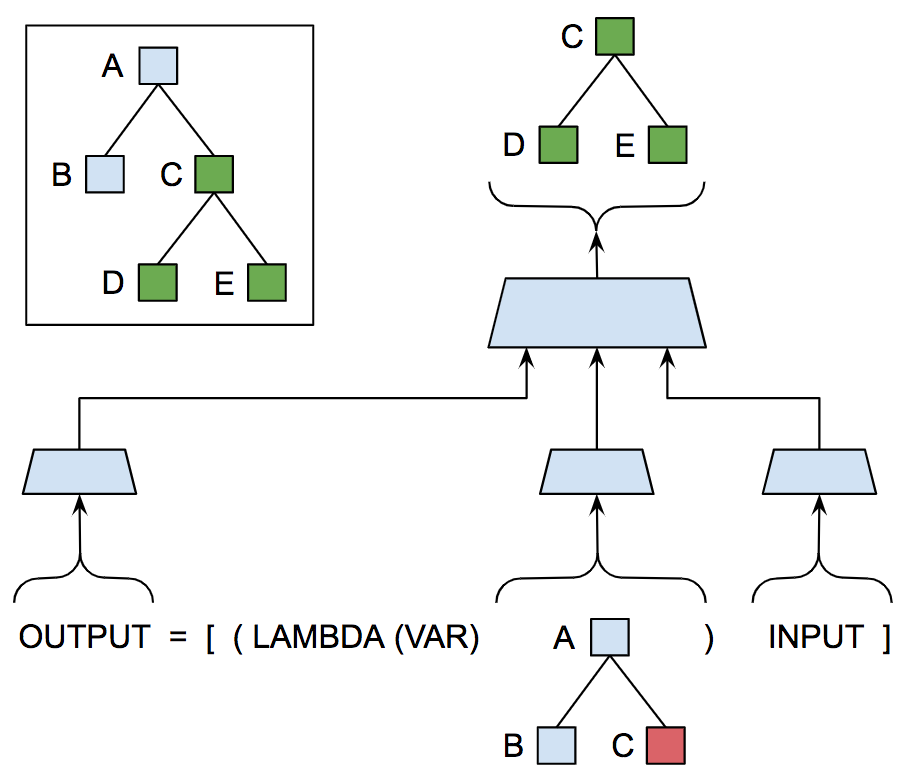} 
  \end{center}
  \caption{This slide illustrates how we make use of input / output pairs as program invariants to narrow search for the next statement in the evolving target program. At any given moment the call stack contains the trace of a single conditioned path through the developing program. A single path is unlikely to provide sufficient information to account for the constraints implicit in all of the sample input / output pairs and so we intend to use a limited lookahead planning system to sample multiple execution traces in order to inform the prediction of the next program statement. 
These so-called imagination-augmented agents implement a novel architecture for reinforcement learning that balances exploration and exploitation using imperfect models to generate trajectories from some initial state using actions sampled from a rollout policy~\cite{PascanuetalCoRR-17,WeberetalCoRR-17,HamricketalCoRR-17,GuezetalCoRR-18}. These trajectories are then combined and fed to an output policy along with the action proposed by a model-free policy to make better decisions. There are related reinforcement learning architectures that perform Monte Carlo Markov chain search to apply and collect the constraints from multiple input / output pairs.}
  \label{fig_Differentiable_Program_Emulation}
\end{figure}


The Differentiable Neural Program (DNP) representation and associated NTM controller for managing the call stack and single-stepping through such programs allow us to exploit the advantages of distributed vector representations to predict the next statement in a program under construction. This model makes it easy to take advantage of supplied natural language descriptions and example input / output pairs plus incorporate semantic information in the form of execution traces generated by utilizing the FIDE to evaluate each statement and encoding information about local variables on the stack. 



Graph Networks is a neural network framework for constructing, modifying and performing inference on differentiable encodings of graphical structures. Battaglia~\etal{}~\cite{BattagliaetalCoRR-18} describe Graph Networks as a "new building block for the AI toolkit with a strong relational inductive bias the {\it{graph network}}, which generalizes and extends various approaches for neural networks that operate on graphs" by constraining the rules governing the composition of entities and their relationships. In principle, graph networks could be used as an alternative to DNPs.

Li~\etal{}~\cite{LietalICLR-18} describe a closely related model they refer to as a Gated Graph Sequence Neural Network (GGS-NN) that operates on graph networks to produce sequences from graph-structured input. Johnson~\cite{JohnsonICLR-17} introduced the Gated Graph Transformer Neural Network (GGT-NN), an extension of GGS-NNs that uses graph-structured data as an intermediate representation. The model can learn to construct and modify graphs in sophisticated ways based on textual input, and also to use the graphs to produce a variety of outputs. The Graph Network (GN) Block described in Section~3.2 of Battaglia~\etal{}~\cite{BattagliaetalCoRR-18} provides a similar set of capabilities.

\begin{figure}
  \begin{center} 
    \includegraphics[width=200pt]{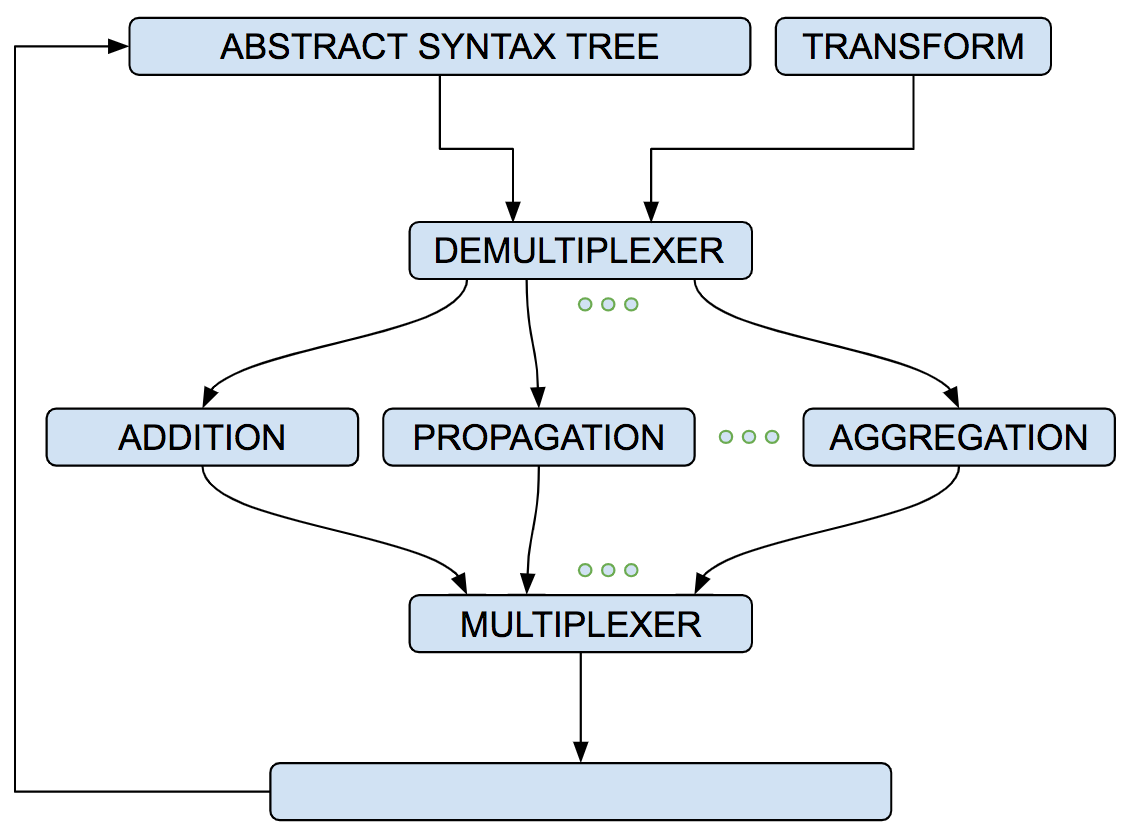} 
  \end{center}
  \caption{The above graphic depicts a utility module that takes a graph in the Graph Networks representation and a command corresponding to one of the transformations described in~\cite{JohnsonICLR-17}, carries out the indicated transformation and produces the transformed graph in a recurrent output layer. See the definition of Graph Network Block in Section~3.2 of Battaglia~\etal{}~\cite{BattagliaetalCoRR-18} for an alternative formulation.}
  \label{fig_Graph_Nets_Transformer_Utility}
\end{figure}

The network shown in Figure~\ref{fig_Graph_Nets_Transformer_Utility} demonstrates how to package the five general transformations described in Johnson's paper to provide a Swiss-army-knife utility that can be used to manipulate abstract syntax trees in code synthesis simplifying the construction of differentiable neural programs introduced earlier. This graph-networks utility could be integrated into a reinforcement-learning code synthesis module that would learn how to repair programs or perform other forms of synthesis by learning how to predict the best alterations on the program under construction. The Graph Network Block provides many of the same operations.


The imagination-based planning (IBP) for reinforcement learning framework~\cite{PascanuetalCoRR-17} serves as an example for how the code synthesis module might be implemented. The IBP architecture combines three separate adaptive components: (a) the {\tt{CONTROLLER}} + {\tt{MEMORY}} system which maps a state $s;\in{};S$ and history $h;\in{};H$ to an action $a;\in{};A$; (b) the {\tt{MANAGER}} maps a history $h;\in{};H$ to a route $u;\in{};U$ that determines whether the system performs an action in the {\tt{COMPUTE}} environment, e.g., single-step the program in the FIDE, or performs an imagination step, e.g., generates a proposal for modifying the existing code under construction; the {\tt{IMAGINATION MODEL}} is a form of dynamical systems model that maps a pair consisting of a state $s;\in{};S$ and an action $a;\in{};A$ to an imagined next state $s';\in{};S$ and scalar-valued reward $r;\in{};R$.

\begin{figure}
  \begin{center} 
    \includegraphics[width=175pt]{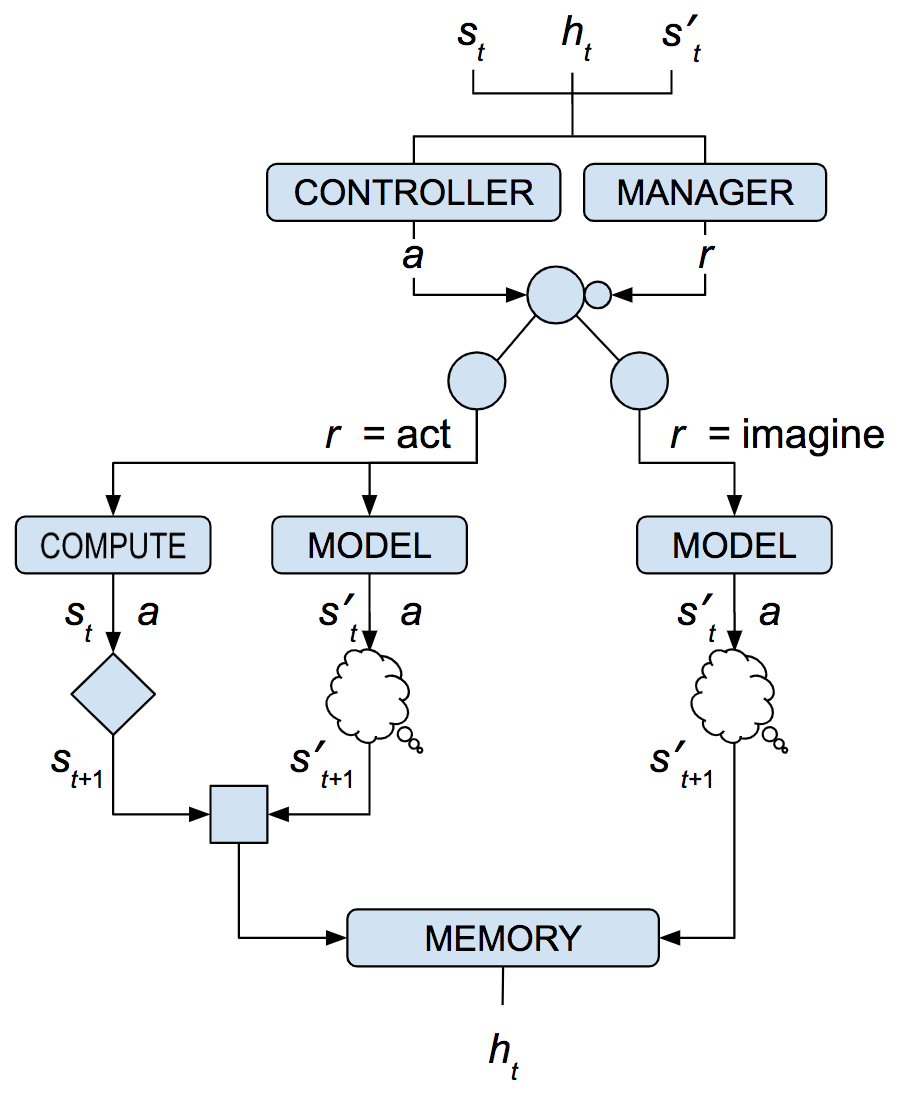} 
  \end{center}
  \caption{The above graphic illustrates how we might adapt the imagination-based planning (IBP) for reinforcement learning framework~\cite{PascanuetalCoRR-17} for use as the core of the apprentice code synthesis module. Actions in this case correspond to transformations of the program under development. States incorporate the history of the evolving partial program. Imagination consists of exploring sequences of program transformations.}
  \label{Graph_Nets_Imagination_Coding}
\end{figure}

The {\tt{IMAGINATION MODEL}} is implemented as an interaction network~\cite{BattagliaetalNIPS-16} that could also be represented using the graph-networks framework. The three components are trained by three distinct, concurrent, on-policy training loops. The IBP framework shown in Figure~\ref{Graph_Nets_Imagination_Coding} allows code synthesis to alternate between exploiting by modifying and running code, and exploring by using the model to investigate and analyze what would happen if you actually did act. The {\tt{MANAGER}} chooses whether to execute a command or predict (imagine) its result and can generate any number of trajectories to produce a tree $h_t$ of imagined results. The {\tt{CONTROLLER}} takes this tree plus the compiled history and chooses an action (command) to carry out in the FIDE.


Context is everything in language and problem solving. When we converse with someone or read a book we keep in mind what was said or written previously. When we attempt to understand what was said in a conversation or formulate what to say next we draw upon our short-term memories of earlier mentioned people and events, but we also draw upon our long-term episodic memories involving the people, places and events related to those explicitly mentioned in the conversation. In solving complex design problems, it is often necessary to keep in mind a large number of specific facts about the different components that go into the design as well as general knowledge pertaining to how those components might be adapted and assembled to produce the final product. 

Much of a programmer's procedural knowledge about how to write code is baked into various cognitive subroutines that can be executed with minimal thinking. For example, writing a simple {\tt{FOR}} loop in Python to iterate through a list is effortless for an experienced Python programmer, but may require careful thought for an analogous code block in a less familiar programming language like C++. In thinking about how the apprentice's knowledge of programming is organized in memory, routine tasks would likely be baked into value functions trained by reinforcement learning. When faced with a new challenge involving unfamiliar concepts or seldom used syntax, we often draw upon less structured knowledge stored in episodic memory. The apprentice uses this same strategy.

The neural network architecture for managing dialogue and writing code involves encoder-decoder pairs comprised of gated recurrent networks that are augmented with attention networks. We'll focus on dialogue to illustrate how context is handled in the process of ingesting (encoding) fragments of an ongoing conversation and generating (decoding) appropriate responses, but the basic architecture is similar for ingesting fragments of code and generating modified fragments that more closely match a specification. The basic architecture employs three attention networks, each of which is associated with a separate encoder network specialized to handle a different type of context. The outputs of the three attention networks are combined and then fed to a single decoder.

The (user response) encoder ingests the most recent utterance produced by the programmer and corresponds to the encoder associated with the encoder-decoder architectures used in machine translation and dialogue management. The (dialogue context) encoder ingests the $N$ words prior to the last utterance. The (episodic memory) encoder ingests older dialogue selected from episodic memory. The attentional machinery responsible for the selection and active maintenance of relevant circuits in the global workspace (GWS) will likely notice and attend to every utterance produced by the programmer. Attentional focus and active maintenance of such circuits in the GWS will result in the corresponding thought vector added to NTM the partition responsible for short-term memory.  

\begin{figure}
  \begin{center} 
    \includegraphics[width=475pt]{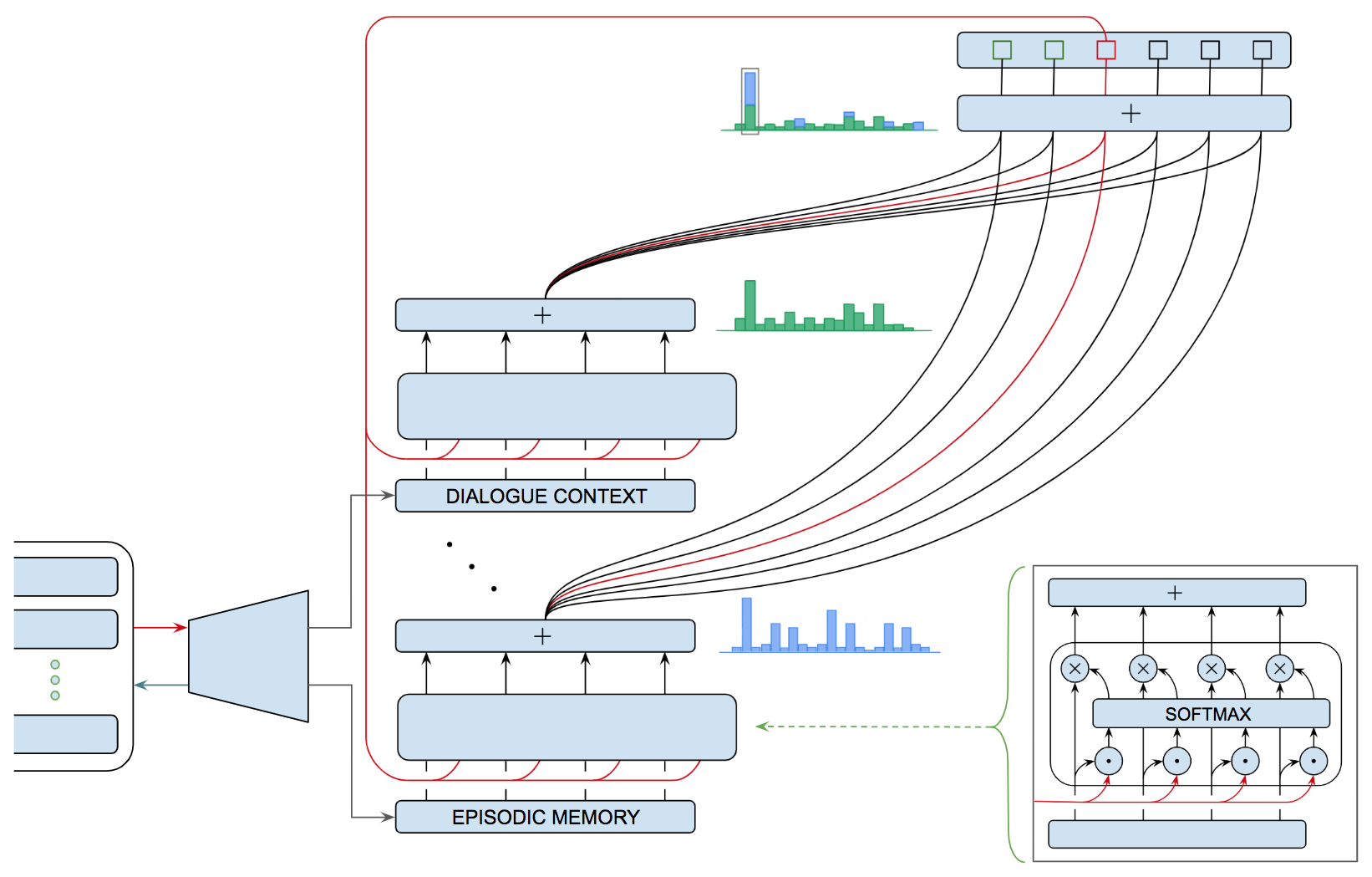} 
  \end{center}
  \caption{In the programmer's assistant, the dialogue management and program-transformation systems are implemented using encoder-decoder sequence-to-sequence networks with attention. We adapt the pointer-generator network model developed by See~\etal{}~\cite{SeeetalACL-17} to combine and bring to bear contextual information from multiple sources including short- and long-term memory systems implemented as Neural Turing Machines as summarized in Figures~\ref{fig_Global_Workspace_Episodic_Memory} and~\ref{fig_High_Level_Assistant_Architecture}. This graphic illustrates two out of the three contextual sources of information employed by the apprentice. Each source is encoded separately, the relevance of its constituent elements represented as a probability distribution and resulting distributions combined to guide the decoder in generating output.}
  \label{Context_Episodic_Memory_Dialog}
\end{figure}

The controller for the NTM partition responsible for short-term (active) memory then generates keys from the newly added thought vectors and transmits these keys to the controller of the NTM partition responsible for long-term (episodic) memory. The episodic memory controller uses these keys to select episodic memories relevant to the current discourse, combining the selected memories into a fixed-length composite thought vector that serves as input for the corresponding encoder. Figure~\ref{Context_Episodic_Memory_Dialog} depicts the basic architecture showing only two of the three encoders and their associated attention networks, illustrating how the outputs of the attention networks are combined prior to being used by the decoder to generate the next word or words in the assistant's next utterance. 


One step of the decoder could add zero, one, or more words, i.e., a phrase, to the current utterance under construction. Memories \emdash{} both short- and long-term \emdash{} are in the form of thought vectors or word sequences that could be used to reconstruct the original thought vectors for embedding or constructing composites by adding context or conditioning to emphasize relevant dimensions. The dialogue manager \emdash{} a $Q$-function network trained by reinforcement learning \emdash{} can also choose not to respond at all or could respond at some length perhaps incorporating references to code, explanations for design choices and demonstrations showing the results of executing code in the IDE.

To control generation, we adapt the pointer-generator network framework developed by See~\etal{} for document summarization~\cite{SeeetalACL-17}. In the standard sequence-to-sequence machine-translation model a weighted average of encoder states becomes the decoder state and attention is just the distribution of weights. In See~\etal{} attention is simpler: instead of weighting input elements, it points at them probabilistically. It isn't necessary to use all the pointers; such networks can mark excerpts by pointing to their start and end constituents. We apply their approach here to digest and integrate contextual information originating from multiple sources.

In humans, memory formation and consolidation involves several systems, multiple stages and can span hours, weeks or months depending on the stage and associated neural circuitry. Our primary interest relates to the earliest stages of memory formation and role of the hippocampus and entorhinal region of the frontal cortex along with several ancillary subcortical circuits including the basal ganglia (BG). Influenced by the work of O'Reilly and Frank{}~\cite{OReillyandFrankNC-06}, we focus on the function of the dentate gyrus (DG) in the hippocampal formation and encode thought vectors using a sparse, invertible mapping thereby providing a high degree of pattern separation in encoding new information while avoiding interference with existing memories. 

We finesse the details of what gets stored and when by simply storing everything. We could store the sparse representation provided by the DG, but prefer to use this probe as the key in a key-value pair in the NTM partition dedicated to episodic memory and store the raw data as the value. This means we have to reconstruct the original encoding produced when initially ingesting the text of an utterance. This is preferable for two reasons: (i) we need the words \emdash{} or tokens of an abstract syntax tree in the case of ingesting code fragments \emdash{} in order for the decoder to generate the apprentice's response, and (ii) the embeddings of the symbolic entities that constitute their meaning are likely to drift during ongoing training.


%% file: inputs/B_Bootstrapped.tex

\section{Bootstrapped Linguistic Competence}


In this appendix, we consider how we might design an end-to-end training protocol for bootstrapping a variant of the programmer's apprentice application. We begin with the analogous stages in early child development. Each of the following four stages is briefly introduced with additional technical details provided in the accompanying footnotes:
\begin{itemize}
\item Basic cognitive bootstrapping and linguistic grounding\endnote{\input{./inputs/footnotes/A_grounding.tex}}: 
  \begin{itemize}
  \item modeling language: statistical $n$-gram language model trained on programming corpus;
  \item hierarchical planning: automated tutor generates lessons using curriculum training;
  \end{itemize}
\item Simple interactive behavior for signaling and editing\endnote{\input{./inputs/footnotes/B_signaling.tex}}: 
  \begin{itemize}
  \item following instruction: learning to carry out simple plans one instruction at a time;
  \item explaining behavior: providing short explanations of behavior, acknowledging failure;
  \end{itemize}
\item Mixed dialogue interleaving instruction and mirroring\endnote{\input{./inputs/footnotes/C_mirroring.tex}}: 
  \begin{itemize}
  \item classifying intention: learning to categorize tasks and summarize intentions to act;
  \item confirming comprehension: conveying practical understanding of specific instructions;
  \end{itemize}
\item Composite behaviors corresponding to simple repairs\endnote{\input{./inputs/footnotes/D_combining.tex}}: 
  \begin{itemize}
  \item executing complex plans: generating and executing multi-step plans with contingencies;
  \item recovering from failure: backtracking, recovering, retracting steps on failed branches;
  \end{itemize}
\end{itemize}


%% file: inputs/footnotes/A_grounding.tex
{\bf{Bootstrapping the programmer's apprentice: Basic cognitive bootstrapping and linguistic grounding:}}\\

The programmer's assistant agent is designed to distinguish between three voices: the voice of the programmer, the voice of the assistant's automated tutor and its own voice. We could have provided an audio track to distinguish these voices, but since there only these three and the overall system can determine when any one of them is speaking, the system simply adds a few bits to each utterance as a proxy for an audio signature allowing the assistant to make such distinctions for itself. When required, we use the same signature to indicate which of the three speakers is responsible for changes to the shared input and output associated with the fully instrumented IDE henceforth abbreviated as FIDE \emdash{} pronounced "/fee/'-/day/", from the Latin meaning: (i) trust, (ii) credit, (iii) fidelity, (iv) honesty. It will also prove useful to further distinguish the voice of the assistant as being in one of two modes: {\it{private}}, engaging in so-called {\it{inner speech}} that is not voiced aloud, and {\it{public}}, meaning spoken out loud for the explicit purpose of communicating with the programmer. We borrow the basic framework for modeling other agents and simple theory-of-mind from Rabinowitz~\etal{}~\cite{RabinowitzetalCoRR-18}.

The bootstrap statistical language model consists of an $n$-gram embedding trained on large general-text language corpus augmented with programming and software-engineering related text drawn from online forums and transcripts of pair-programming dialog. For the time being, we will not pursue the option of trying to acquire a large enough dialog corpus to train an encoder-decoder LSTM/GRU dialog manager / conversational model~\cite{VinyalsandLeICML-15}. In the initial prototype, natural language generation (NLG) output for the automated tutor and assistant will be handled using hierarchical planning technology leveraging ideas developed in the CMU RavenClaw dialogue management system~\cite{BohusandRudnickyCS-09}, but we have plans to explore hybrid natural language generation by combining hard-coded Python dialog agents corresponding to hierarchical task networks and differentiable dialogic encoder-decoder thought-cloud generators using a variant of pointer-generator networks as described by See~\etal{}~\cite{SeeetalACL-17}.

Both the tutor and assistant NLG subsystems will rely on a base-level collection of plans \emdash{} hierarchical task network (HTN) \emdash{} that we employ in several contexts plus a set of specialized plans \emdash{} an HTN subnetwork \emdash{} specific to each subsystem. At any given moment in time, a meta control system~\cite{HamricketalCoRR-17} in concert with a reinforcement-learning-trained policy determines the curricular goal constraining the tutor's choice of specific lesson is implemented using a variant of the scheduled auxiliary control paradigm described by Riedmiller~\etal{}~\cite{RiedmilleretalCoRR-18}. Having selected a subset of lessons relevant to the current curricular goal, the meta-controller cedes control to the tutor which selects a specific lesson and a suitable plan to oversee interaction with the agent over the course of the lesson.

Most lessons will require a combination of spoken dialogue and interactive signaling that may include both the agent and the tutor pointing, highlighting, performing edits and controlling the FIDE by executing code and using developer tools like the debugger to change state, set break points and single step the interpreter, but we're getting ahead of ourselves. The curriculum for mastering the basic referential modes is divided into three levels of mastery in keeping with Terrence Deacon's description~\cite{Deacon1998symbolic} and Charles Sanders Peirce's (semiotic) {\it{theory of signs}}. The tutor will start at the most basic level, continually evaluating performance to determine when it is time to graduate to the next level or when it is appropriate to revert to an earlier level to provide additional training in order to master the less demanding modes of reference.\\


%% file: inputs/footnotes/B_signaling.tex
{\bf{Bootstrapping the programmer's apprentice: Simple interactive behavior for signaling and editing:}}\\

In the first stage of bootstrapping, the assistant's automated tutor engages in an analog of the sort of simple signaling and reinforcement that a mother might engage in with her baby in order to encourage the infant to begin taking notice of its environment and participating in the simplest forms of communication. The basic exchange goes something like: the mother draws the baby's attention to something and the baby acknowledges by making some sound or movement. This early step requires that the baby can direct its gaze and attend to changes in its visual field.

In the case of the assistant, the relevant changes would correspond to changes in FIDE or the shared browser window, pointing would be accomplished by altering the contents of FIDE buffers or modifying HTML. Since the assistant has an innate capability to parse language into sequences of words, the tutor can preface each lesson with short verbal lesson summary, e.g., "the variable '{\tt{foo}}'", "the underlined variable", "the highlighted assignment statement", "the expression highlighted in blue". The implicit curriculum followed by the tutor would systematically graduate to more complicated language for specifying referents, e.g., "the {\it{body}} of the '{\tt{for}}' loop", "the '{\tt{else}}' {\it{clause}} in the '{\tt{conditional}} statement", "the {\it{scope}} of the variable '{\tt{counter}}'", "the expression on the {\it{right-hand side}} of the first assignment statement".

The goal of the bootstrap tutor is to eventually graduate to simple substitution and repair activities requiring a combination of basic attention, signaling, requesting feedback and simple edits, e.g., "highlight the scope of the variable shown in red", "change the name of the function to be "{\tt{Increment\_Counter}}", "insert a "{\tt{for}}" loop with an iterator over the items in the "{\tt{bucket}}" list", "delete the next two expressions", with the length and complexity of the specification gradually increasing until the apprentice is capable of handling code changes that involve multiple goals and dozens of intermediate steps, e.g., "delete the variable "{\tt{Interrupt\_Flag}}" from the parameter list of the function declaration and eliminate all of the expressions that refer to the variable within the scope of the function definition".

Note the importance of an attentional system that can notice changes in the integrated development environment and shared browser window, the ability to use recency to help resolve ambiguities, and emphasize basic signals that require noticing changes in the IDE and acknowledging that these changes were made as a means of signaling expectations relevant to the ongoing conversation between the programmer and the apprentice. These are certainly subtleties that will have to be introduced gradually into the curricular repertoire as the apprentice gains experience. We are depending on employing a variant of Riedmiller~\etal{} that will enable us to employ the FIDE to gamify the process by evaluating progress at different levels using a combination of general extrinsic reward and policy-specific intrinsic motivations to guide action selection~\cite{RiedmilleretalCoRR-18}.

Randall O'Reilly mentioned in his class presentation the idea that natural language might play an important role in human brains as an intra-cortical lingua franca. Given that one of the primary roles language serves is to serialize thought thereby facilitating serial computation with all of its advantages in terms of logical precision and combinatorial expression, projecting a distributed connectionist representation through some sort of auto encoder bottleneck might gain some advantage in combining aspects of symbolic and connectionist architectures. This also relates to O'Reilly's discussion of the hippocampal system and in particular the processing performed by the dentate gyrus and hippocampal areas CA1 in CA2 in generating a sparse representation that enables rapid binding of arbitrary informational states and facilitates encoding and retrieving of episodic memory in the entorhinal cortex.\\


%% file: inputs/footnotes/C_mirroring.tex
{\bf{Bootstrapping the programmer's apprentice: Mixed dialogue interleaving instruction and mirroring:}}\\

Every utterance, whether generated by the programmer or the apprentice's tutor or generated by the apprentice either intended for the programmer or {\it{sotto voce}} for its internal record, has potential future value and hence it makes sense to record that utterance along with any context that might help to realize that potential at a later point in time. Endel Tulving coined the phrase {\it{episodic memory}} to refer to this sort of memory. We'll forgo discussion of other types of memory for the time being and focus on what the apprentice will need to remember in order take advantage of its past experience. 

Here is the simplest, stripped-to-its-most-basic-elements scenario outlined in the class notes: (a) the apprentice performs a sequence of steps that effect a repair on a code fragment, (b) this experience is recorded in a sequence of tuples of the form $(s_{t}, a_{t}, r_{t}, s_{t+1})$ and consolidated in episodic memory, (c) at a subsequent time, days or weeks later, the apprentice recognizes a similar situation and realizes an opportunity to exercise what was learned in the earlier episode, and (d) a suitably adapted repair is applied in the present circumstances and incorporated into a more general policy so that it can be applied in wider range circumstances.

The succinct notation doesn't reveal any hint of the complexity and subtlety of the question. What were the (prior) circumstances \emdash{} $s_{t}$? What was thought, said and done to plan, prepare and take action \emdash{} $a_{t}$? What were the (posterior) consequences \emdash{} $r_{t}$ and $s_{t+1}$? We can't simply record the entire neural state vector. We could, however, plausibly record the information temporarily stored in working memory since this is the only information that could have played any substantive role \emdash{} for better or worse \emdash{} in guiding executive function. 

We can't store everything and then carefully pick through the pile looking for what might have made a difference, but we can do something almost as useful. We can propagate the reward gradient back through the value- / $Q$-function and then further back through the activated circuits in working memory that were used to select $a_{i}$ and adjust their weights accordingly. The objective in this case being to optimize the $Q$-function by predicting the state variables that it needs in order to make an accurate prediction of the value of applying action $a_{t}$ in $s_{t}$ as described in Wayne~\etal{}~\cite{WayneetalCoRR-18}.

Often the problem can be described as a simple Markov process and the state represented as a vector comprising of a finite number of state variables, $s_t = \langle{} \sigma{}_0, \sigma{}_1, \sigma{}_2, \sigma{}_3, \sigma{}_4, \sigma{}_5, \sigma{}_6, \sigma{}_7, \rangle{}$, with the implicit assumption that the process is {\it{fully observable}}. More generally, the Markov property still holds, but the state is only {\it{partially observable}} resulting in a much harder class of decision problem known as a {\urlh{https://en.wikipedia.org/wiki/Partially_observable_Markov_decision_process}{POMDP}}. In some cases, we can finesse the complexity if we can ensure that we can observe the {\it{relevant}} state variables in any given state, e.g., in one set of states it is enough to know one subset of the state variables, $\langle{} \sigma{}_0, \color{red}\sigma{}_1\color{black}, \sigma{}_2, \color{red}\sigma{}_3\color{black}, \sigma{}_4, \sigma{}_5, \sigma{}_6, \color{red}\sigma{}_7\color{black} \rangle{}$, while in another set of states a different subset of state variables suffices, $\langle{} \sigma{}_0, \sigma{}_1, \sigma{}_2, \sigma{}_3, \color{red}\sigma{}_4\color{black}, \sigma{}_5, \color{red}\sigma{}_6\color{black}, \sigma{}_7 \rangle{}$. If you can learn which state variables are required and arrange to observe them, the problem reduces to the fully observed case. 

There's a catch however. The state vector includes state variables that correspond to the observations of external processes that we have little or no direct control over as well as the apprehension of internal processes including the activation of subnetworks. We may need to plan for and carry out the requisite observations to acquire the external process state and perform the requisite computations to produce and then access the resulting internal state information. We also have the ability to perform two fundamentally different types of computation each of which has different strengths and weaknesses that conveniently complement the other.

The mammalian brain is optimized to efficiently perform many computations in parallel; however, for the most part it is not particularly effective dealing with the inconsistencies that arise among those largely independent computations. Rather than relying on estimating and conditioning action selection on internally maintained state variables, most animals rely on environmental cues \emdash{} callsed {\urlh{https://en.wikipedia.org/wiki/Affordance}{affordances}}~\cite{GibsonAFFORDANCES-79} \emdash{} to restrict the space of possible options and simplify action selection. However, complex skills like programming require complex serial computations in order to reconcile and make sense of the contradictory suggestions originating from our mostly parallel computational substrate. 

Conventional reinforcement learning may work for some types of routine programming like writing simple text-processing scripts, but it is not likely to suffice for programs that involve more complex logical, mathematical and algorithmic thinking. The programmer's apprentice project is intended as a playground in which to explore ideas derived from biological systems that might help us chip away at these more difficult problems. For example, the primate brain compensates for the limitations of its largely parallel processing approach to solving problems by using specialized networks in the frontal cortex, thalamus, striatum, and basal ganglia to serialize the computations necessary to perform complex thinking. 

At the very least, it seems reasonable to suggest that we need cognitive machinery that is at least as powerful as the programs we aspire the apprentice to generate~\cite{GallistelandKing2009computational}. We need the neural equivalent of the {{\tt{CONTROL UNIT}}} responsible for maintaining a {\tt{PROGRAM COUNTER}} and the analog of loading instructions and operands into {\tt{REGISTERS}} in the {\tt{ARITHMETIC AND LOGIC UNIT}} and subsequently writing the resulting computed products into other registers or {\tt{RANDOM ACCESS MEMORY}}. These {\it{particular}} features of the von Neumann architecture are not essential \emdash{} what is required is a lingistic foundation that supports a complete story of computation and that is grounded in the detailed \emdash{} almost visceral \emdash{} experience of carrying out computations.

A single $Q$ (value) function encoding a single action-selection policy with fixed finite-discrete or continuous state and action spaces isn't likely to suffice. Supporting compiled subroutines doesn't significantly change the picture. The addition of a meta controller for orchestrating a finite collection of separate, special-purpose policies adds complexity without appreciably adding competence. And simply adding language for describing procedures, composing production rules, and compiling subroutines as a Sapir-Whorf-induced infusion of ontological enhancement is \emdash{} by itself \emdash{} only a distraction. We need an approach that exploits a deeper understanding of the role of language in the modern age \emdash{} a method of using a subset of natural language to describe programs in terms of narratives where executing such a program is tantamount to telling the story. Think about how human cognitive systems encode and serialize remembered stories, about programs as stories drawing on life experience by exploiting the serial nature of episodic memory, and about thought clouds that represent a superposition of eigenstates such that collapsing the wave function yields coherent narrative that serves as a program trace.\\


%% file: inputs/footnotes/D_combining.tex
{\bf{Bootstrapping the programmer's apprentice: Composite behaviors corresponding to simple repairs:}}\\

A {\it{software design pattern}} "is a general, reusable solution to a commonly occurring problem within a given context in software design. It is not a finished design that can be transformed directly into source or machine code. It is a description or template for how to solve a problem that can be used in many different situations. Design patterns are formalized best practices that the programmer can use to solve common problems when designing an application or system". They are typically characterized as belonging to one of three categories: {\it{creational}}, {\it{structural}}, or {\it{behavioral}}.

We would like to believe that such patterns provide clear prescriptions for how to tackle challenging programming problems, but know better. Studying such patterns and analyzing examples of their application to practical problems is an excellent exercise for both computer science students learning to program, and practicing software engineers wanting to improve their skills. That said, these design patterns require considerable effort to master and are well beyond what one might hope to accomplish in bootstrapping basic linguistic and programming skills. Indeed, mastery depends on already knowing \emdash{} at the very least \emdash{} the rudiments of these skills. 


We are willing to concede that mental software is not always expressed in language. For the programmer's apprentice, we are thinking of encoding what is essentially static and syntactic knowledge about programs and programming using four different representations, and what is essentially dynamic and semantic knowledge in a family of structured representations that encode program execution traces of one sort or another. The four static / syntactic representations are summarized as follows:
\begin{itemize}
\item (i) distributed (connectionist) representations of natural language as points in high-dimensional embedding spaces \emdash{} thought clouds;
\item (ii) natural language transcripts of dialogical utterances / interlocutionary acts encoded as lexical token streams \emdash{} word sequences;
\item (iii) programs in the target programming language represented as structured objects corresponding to augmented {\it{abstract syntax trees}} (ASTs)\emdash{} the augmentations correspond to edges representing procedure calls, iteration and recursion resulting in directed acyclic graphs;
\item (iv) hierarchical plans corresponding to subnetworks of {\it{hierarchical task networks}} (HTNs) or, if you like, the implied representation of hierarchical plans encoded in value iteration networks~\cite{TamaretalNIPS-16} and goal-based policies~\cite{GroshevetalCoRR-17}. We're also thinking about encoding HTNs as policies using a variation on the idea of {\it{options}}~\cite{SuttonetalAIJ-99} as described in Riedmiller~\etal{}~\cite{RiedmilleretalCoRR-18}; 
\end{itemize}

The first entry (i) is somewhat misleading in that any one of the remaining three (ii-iv) can be represented as a point / thought cloud using an appropriate embedding method. Thought clouds are the Swiss Army knife of distributed codes. They represent a (constrained) superposition of possibilities allowing us to convert large corpora of serialized structures into point clouds that enable massively parallel search, and subsequently allow us to collapse the wave function, as it were, to read off solutions by re-serializing the distributed encoding of constraints that result from conducting such parallel searches.

We propose to develop encoders and decoders to translate between (serial) representations (ii-iv) where only a subset of conversions are possible or desirable given the expressivity of the underlying representation language. We imagine autoencoders with an {\it{information bottleneck}} that take embeddings of natural language descriptions as input and produces an equivalent HTN representation, combining a mixture of (executable) interlocutory and code synthesis tasks~\cite{AlemietalCoRR-16b,TishbyandZaslavskyCoRR-15}. The interlocutory tasks generate explanations and produce comments and specifications. The code-synthesis tasks serve to generate, repair, debug and test code represented in the FIDE.

Separately encoded embeddings will tend to evolve independently, frustrating attempts to combine them into composite representations that allow powerful means of abstraction. The hope is that we can use natural language as a {\urlh{https://en.wikipedia.org/wiki/Lingua_franca}{lingua franca}} \emdash{} a "bridge" language \emdash{} to coerce agreement among disparate representations by forcing them to cohere along shared, possibly refactored dimensions in much the same way that {\it{trade languages}} serve as an expeditious means of exchanging information between scientists and engineers working in different disciplines or scholars who do not share a native language or dialect.\\


%% file: paper.bbl
\begin{thebibliography}{100}

\bibitem{AbolafiaetalCoRR-18}
Daniel~A. Abolafia, Mohammad Norouzi, and Quoc~V. Le.
\newblock Neural program synthesis with priority queue training.
\newblock {\em CoRR}, arXiv:1801.03526, 2018.

\bibitem{AlemietalCoRR-16b}
Alexander~A. Alemi, Ian Fischer, Joshua~V. Dillon, and Kevin Murphy.
\newblock Deep variational information bottleneck.
\newblock {\em CoRR}, arXiv:1612.00410, 2016.

\bibitem{BaetalCoRR-16}
Jimmy Ba, Geoffrey Hinton, Volodymyr Mnih, Joel~Z. Leibo, and Catalin Ionescu.
\newblock Using fast weights to attend to the recent past.
\newblock {\em CoRR}, arXiv:1610.06258, 2016.

\bibitem{Baars1988}
B.~J. Baars.
\newblock {\em A cognitive theory of consciousness}.
\newblock Cambridge University Press, New York, NY, 1988.

\bibitem{Ballard2015}
Dana~H. Ballard.
\newblock {\em Brain Computation as Hierarchical Abstraction}.
\newblock Computational Neuroscience Series. MIT Press, 2015.

\bibitem{BattagliaetalNIPS-16}
Peter Battaglia, Razvan Pascanu, Matthew Lai, Danilo~Jimenez Rezende, and Koray
  kavukcuoglu.
\newblock Interaction networks for learning about objects, relations and
  physics.
\newblock In {\em Proceedings of the 30th International Conference on Neural
  Information Processing Systems}, pages 4509-4517. Curran Associates Inc.,
  2016.

\bibitem{BattagliaetalCoRR-18}
Peter~W. Battaglia, Jessica~B. Hamrick, Victor Bapst, Alvaro Sanchez-Gonzalez,
  Vinicius Zambaldi, Mateusz Malinowski, Andrea Tacchetti, David Raposo, Adam
  Santoro, Ryan Faulkner, Caglar Gulcehre, Francis Song, Andrew Ballard, Justin
  Gilmer, George Dahl, Ashish Vaswani, Kelsey Allen, Charles Nash, Victoria
  Langston, Chris Dyer, Nicolas Heess, Daan Wierstra, Pushmeet Kohli, Matt
  Botvinick, Oriol Vinyals, Yujia Li, and Razvan Pascanu.
\newblock Relational inductive biases, deep learning, and graph networks.
\newblock {\em CoRR}, arXiv:1806.01261, 2018.

\bibitem{BengioCoRR-17}
Yoshua Bengio.
\newblock The consciousness prior.
\newblock {\em CoRR}, arXiv:1709.08568, 2017.

\bibitem{BengioetalICML-09}
Yoshua Bengio, J{\'e}r\^{o}me Louradour, Ronan Collobert, and Jason Weston.
\newblock Curriculum learning.
\newblock In {\em Proceedings of the 26th Annual International Conference on
  Machine Learning}, pages 41-48, New York, NY, USA, 2009. ACM.

\bibitem{BertotandCasteranCOQ-04}
Yves Bertot and Pierre Cast\'{e}ran.
\newblock {\em Interactive Theorem Proving and Program Development: Coq'Art The
  Calculus of Inductive Constructions}.
\newblock Springer, New York, 2010.

\bibitem{BhoopchandetalICLR-17}
Avishkar Bhoopchand, Tim Rockt\"{a}schel, Earl~T. Barr, and Sebastian Riedel.
\newblock Learning python code suggestion with a sparse pointer network.
\newblock In {\em International Conference on Learning Representations}, volume
  arXiv:/1611.08307, 2017.

\bibitem{BinderandDesaiTiCS-11}
Jeffrey~R. Binder and Rutvik~H. Desai.
\newblock The neurobiology of semantic memory.
\newblock {\em Trends in Cognitive Science}, 15:527-536, 2011.

\bibitem{BohusPhD-07}
Dan Bohus.
\newblock {\em Error Awareness and Recovery in Conversational Spoken Language
  Interfaces}.
\newblock PhD thesis, Carnegie Mellon University, 2007.

\bibitem{BohusandRudnickyCSL-09}
Dan Bohus and Alexander~I. Rudnicky.
\newblock {The {RavenClaw} dialogue management framework: architecture and
  systems}.
\newblock {\em Computer Speech \& Language}, 23:332-361, 2009.

\bibitem{BohusandRudnickyCS-09}
Dan Bohus and Alexander~I. Rudnicky.
\newblock {The RavenClaw Dialogue Management Framework: Architecture and
  Systems}.
\newblock {\em Computer Speech and Language}, 23:332-361, 2009.

\bibitem{BrodieetalICAC-05}
M.~Brodie, Sheng Ma, G.~Lohman, L.~Mignet, N.~Modani, M.~Wilding, J.~Champlin,
  and P.~Sohn.
\newblock Quickly finding known software problems via automated symptom
  matching.
\newblock In {\em Proceedings of the Second International Conference on
  Autonomic Computing}. ACM, New York, NY, USA, 2005.

\bibitem{CaietalICLR-17}
Jonathon Cai, Richard Shin, and Dawn Song.
\newblock Making neural programming architectures generalize via recursion.
\newblock {\em CoRR}, arXiv:1704.06611, 2017.

\bibitem{ChaterandChristiansenHLB-11}
Nick Chater and Morten~H. Christiansen.
\newblock A solution to the logical problem of language evolution: language as
  an adaptation to the human brain.
\newblock In Kathleen~R. Gibson and Maggie Tallerman, editors, {\em The Oxford
  Handbook of Language Evolution}. Oxford University Press, 2011.

\bibitem{ChaterandChristiansenCOiBS-18}
Nick Chater and Morten~H. Christiansen.
\newblock Language acquisition as skill learning.
\newblock {\em Current Opinion in Behavioral Sciences}, 2018.

\bibitem{ChateretalJML-16}
Nick Chater, Stewart~M. McCauley, and Morten~H. Christiansen.
\newblock Language as skill: Intertwining comprehension and production.
\newblock {\em Journal of Memory and Language}, 89:244-254, 2016.

\bibitem{ChenetalICLR-18b}
Xinyun Chen, Chang Liu, and Dawn Song.
\newblock Tree-to-tree neural networks for program translation.
\newblock {\em CoRR}, arXiv:1802.03691, 2018.

\bibitem{ChistyakovetalICLR-17}
Alexander Chistyakov, Ekaterina Lobacheva, Arseny Kuznetsov, and Alexey
  Romanenko.
\newblock Semantic embeddings for program behaviour patterns.
\newblock In {\em ICLR Workshop}, 2017.

\bibitem{DasetalCVPR-17}
Abhishek Das, Satwik Kottur, Khushi Gupta, Avi Singh, Deshraj Yadav,
  Jos\'e~M.F. Moura, Devi Parikh, and Dhruv Batra.
\newblock {V}isual {D}ialog.
\newblock In {\em Proceedings of the IEEE Conference on Computer Vision and
  Pattern Recognition (CVPR)}, 2017.

\bibitem{Deacon1998symbolic}
Terrence~W. Deacon.
\newblock {\em The Symbolic Species: The Co-evolution of Language and the
  Brain}.
\newblock W. W. Norton, 1998.

\bibitem{DeanAAAI-05}
Thomas Dean.
\newblock A computational model of the cerebral cortex.
\newblock In {\em Proceedings of AAAI-05}, pages 938-943, Cambridge,
  Massachusetts, 2005. MIT Press.

\bibitem{DeanAMAI-06}
Thomas Dean.
\newblock Learning invariant features using inertial priors.
\newblock {\em Annals of Mathematics and Artificial Intelligence}, 47:223-250,
  2006.

\bibitem{Dehaene2014}
Stanislas Dehaene.
\newblock {\em Consciousness and the Brain: Deciphering How the Brain Codes Our
  Thoughts}.
\newblock Viking Press, 2014.

\bibitem{DehaeneetalPNAS-98}
Stanislas Dehaene, Michel Kerszberg, and Jean-Pierre Changeux.
\newblock A neuronal model of a global workspace in effortful cognitive tasks.
\newblock {\em Proceedings of the National Academy of Sciences},
  95:14529-14534, 1998.

\bibitem{DehaeneetalSCIENCE-17}
Stanislas Dehaene, Hakwan Lau, and Sid Kouider.
\newblock What is consciousness, and could machines have it?
\newblock {\em Science}, 358(6362):486-492, 2017.

\bibitem{DevlinetalICLR-18}
Jacob Devlin, Jonathan Uesato, Rishabh Singh, and Pushmeet Kohli.
\newblock Semantic code repair using neuro-symbolic transformation networks.
\newblock In {\em International Conference on Learning Representations}, 2018.

\bibitem{DonnarummaetalIJNS-15}
F.~Donnarumma, R.~Prevete, F.~Chersi, and G.~Pezzulo.
\newblock A programmer-interpreter neural network architecture for prefrontal
  cognitive control.
\newblock {\em International Journal Neural Systems}, 25(6):1550017, 2015.

\bibitem{FanetalJCN-17}
D.~Fan, Q.~Wang, J.~Su, and H.~Xi.
\newblock Stimulus-induced transitions between spike-wave discharges and
  spindles with the modulation of thalamic reticular nucleus.
\newblock {\em Journal of Computational Neuroscience}, 43(3):203-225, 2017.

\bibitem{FinnetalCoRR-17}
Chelsea Finn, Pieter Abbeel, and Sergey Levine.
\newblock Model-agnostic meta-learning for fast adaptation of deep networks.
\newblock {\em CoRR}, arXiv:1703.03400, 2017.

\bibitem{FodorandPylyshynCOGNITION-88}
Jerry~A. Fodor and Zenon~W. Pylyshyn.
\newblock Connectionism and cognitive architecture.
\newblock {\em Cognition}, 28(1-2):3-71, 1988.

\bibitem{FoersteretalCoRR-17}
Jakob~N. Foerster, Richard~Y. Chen, Maruan Al{-}Shedivat, Shimon Whiteson,
  Pieter Abbeel, and Igor Mordatch.
\newblock Learning with opponent-learning awareness.
\newblock {\em CoRR}, abs/1709.04326, 2017.

\bibitem{GallistelandKing2009computational}
C.R. Gallistel and A.P. King.
\newblock {\em {Memory and the Computational Brain: Why Cognitive Science will
  Transform Neuroscience}}.
\newblock Wiley, 2009.

\bibitem{GeorgeandHawkinsIJCNN-05}
Dileep George and Jeff Hawkins.
\newblock A hierarchical {B}ayesian model of invariant pattern recognition in
  the visual cortex.
\newblock In {\em Proceedings of the International Joint Conference on Neural
  Networks}, volume~3, pages 1812-1817. IEEE, 2005.

\bibitem{GibsonAFFORDANCES-79}
James~J. Gibson.
\newblock {\em The Ecological Approach to Visual Perception}.
\newblock Houghton Mifflin, Boston, 1979.

\bibitem{CS379C_Final_Project_Gomezetal-18}
Marcus Gomez, Nate Gruver, Michelle Lam, Rohun Saxena, and Lucy Wang.
\newblock Imitation learning for code generation via recurrent state space
  embeddings.
\newblock Representative Final Project Archive for Stanford CS379C in Spring
  2018, 2018.

\bibitem{GravesetalCoRR-14}
Alex Graves, Greg Wayne, and Ivo Danihelka.
\newblock Neural {T}uring machines.
\newblock {\em CoRR}, arXiv:1410.5401, 2014.

\bibitem{GravesetalNATURE-16}
Alex Graves, Greg Wayne, Malcolm Reynolds, Tim Harley, Ivo Danihelka, Agnieszka
  Grabska-Barwi\'{n}ska, Sergio~G\'{o}mez Colmenarejo, Edward Grefenstette,
  Tiago Ramalho, John Agapiou, Adri\`{a}~Puigdom\'{e}nech Badia, Karl~Moritz
  Hermann, Yori Zwols, Georg Ostrovski, Adam Cain, Helen King, Christopher
  Summerfield, Phil Blunsom, Koray Kavukcuoglu, and Demis Hassabis.
\newblock Hybrid computing using a neural network with dynamic external memory.
\newblock {\em Nature}, 538:471-476, 2016.

\bibitem{GrazianoFiRAI-17}
Michael S.~A. Graziano.
\newblock The attention schema theory: A foundation for engineering artificial
  consciousness.
\newblock {\em Frontiers in Robotics and AI}, 4:60, 2017.

\bibitem{GroshevetalCoRR-17}
Edward Groshev, Aviv Tamar, Siddharth Srivastava, and Pieter Abbeel.
\newblock Learning generalized reactive policies using deep neural networks.
\newblock {\em CoRR}, arXiv:1708.07280, 2017.

\bibitem{GuezetalCoRR-18}
Arthur Guez, Th{\'{e}}ophane Weber, Ioannis Antonoglou, Karen Simonyan, Oriol
  Vinyals, Daan Wierstra, R{\'{e}}mi Munos, and David Silver.
\newblock Learning to search with {MCTS}nets.
\newblock {\em CoRR}, arXiv:1802.04697, 2018.

\bibitem{GulwanietalFaTiPL-17}
Sumit Gulwani, Oleksandr Polozov, and Rishabh Singh.
\newblock Program synthesis.
\newblock {\em Foundations and Trends in Programming Languages}, 4(1-2):1-119,
  2017.

\bibitem{GuptaetalPATENT-06}
Rajeev Gupta, Guy~Maring Lohman, Tanveer~Fathima Mahmood, Laurent~Sebastien
  Mignet, Natwar Modani, and Mark~Francis Wilding.
\newblock System and method for matching a plurality of ordered sequences with
  applications to call stack analysis to identify known software problems.
\newblock https://patents.google.com/patent/US7840946, 2006.

\bibitem{HamricketalCoRR-17}
Jessica~B. Hamrick, Andrew~J. Ballard, Razvan Pascanu, Oriol Vinyals, Nicolas
  Heess, and Peter~W. Battaglia.
\newblock Metacontrol for adaptive imagination-based optimization.
\newblock {\em CoRR}, arXiv:1705.02670, 2017.

\bibitem{HassabisandMaguireTiCS-07}
Demis Hassabis and Eleanor~A. Maguire.
\newblock Deconstructing episodic memory with construction.
\newblock {\em Trends in Cognitive Science}, 11:299-306, 2007.

\bibitem{Hawkins04}
Jeff Hawkins and Sandra Blakeslee.
\newblock {\em On Intelligence}.
\newblock Henry Holt and Company, New York, 2004.

\bibitem{HoandErmonCoRR-16}
Jonathan Ho and Stefano Ermon.
\newblock Generative adversarial imitation learning.
\newblock {\em CoRR}, arXiv:1606.03476, 2016.

\bibitem{HubelandWieselJoP-62}
D.~H. Hubel and T.~N Wiesel.
\newblock Receptive fields, binocular interaction and functional architecture
  in the cat's visual cortex.
\newblock {\em Journal of Physiology}, 160:106-154, 1962.

\bibitem{HubelandWieselJoP-68}
D.~H. Hubel and T.~N Wiesel.
\newblock Receptive fields and functional architecture of monkey striate
  cortex.
\newblock {\em Journal of Physiology}, 195:215-243, 1968.

\bibitem{JohnsonICLR-17}
Daniel~D. Johnson.
\newblock Learning graphical state transitions.
\newblock In {\em International Conference on Learning Representations}, 2017.

\bibitem{KavukcuogluetalNIPS-10}
Koray Kavukcuoglu, Pierre Sermanet, Y-Lan Boureau, Karol Gregor, Micha\"{e}l
  Mathieu, and Yann LeCun.
\newblock Learning convolutional feature hierarchies for visual recognition.
\newblock In {\em Proceedings of the 23rd International Conference on Neural
  Information Processing Systems - Volume 1}, pages 1090-1098. Curran
  Associates Inc., 2010.

\bibitem{KingmaandWellingCoRR-13}
Diederik~P. Kingma and Max Welling.
\newblock Auto-encoding variational {B}ayes.
\newblock {\em CoRR}, arXiv:1312.6114, 2013.

\bibitem{KrieteetalPNAS-13}
Trenton Kriete, David~C. Noelle, Jonathan~D. Cohen, and Randall~C. O'Reilly.
\newblock Indirection and symbol-like processing in the prefrontal cortex and
  basal ganglia.
\newblock {\em Proceedings of the National Academy of Sciences}, 2013.

\bibitem{Kurzweil2012}
Ray Kurzweil.
\newblock {\em How to Create a Mind: The Secret of Human Thought Revealed}.
\newblock Viking Press, New York, NY, 2012.

\bibitem{LatchoumaneetalNEURON-17}
C.~V. Latchoumane, H.~V. Ngo, J.~Born, and H.~S. Shin.
\newblock Thalamic {S}pindles {P}romote {M}emory {F}ormation during {S}leep
  through {T}riple {P}hase-{L}ocking of {C}ortical, {T}halamic, and
  {H}ippocampal {R}hythms.
\newblock {\em Neuron}, 95(2):424-435, 2017.

\bibitem{LevyandGoldbergCONIL-14}
Omer Levy and Yoav Goldberg.
\newblock Linguistic regularities in sparse and explicit word representations.
\newblock In {\em Proceedings of the Eighteenth Conference on Computational
  Natural Language Learning}, pages 171-180, Ann Arbor, Michigan, 2014.
  Association for Computational Linguistics.

\bibitem{LietalICLR-18}
Yujia Li, Oriol Vinyals, Chris Dyer, Razvan Pascanu, and Peter Battaglia.
\newblock Learning deep generative models of graphs.
\newblock In {\em International Conference on Learning Representations}, 2018.

\bibitem{CS379C_Final_Project_Luetal-18}
Peter Lu, Sophia Sanchez, Yousef Hindy, Maurice Chiang, and Michael Smith.
\newblock Integrating reinforcement learning agents for error correction in
  abstract syntax trees.
\newblock Representative Final Project Archive for Stanford CS379C in Spring
  2018, 2018.

\bibitem{MerityetalCoRR-16}
Stephen Merity, Caiming Xiong, James Bradbury, and Richard Socher.
\newblock Pointer sentinel mixture models.
\newblock {\em CoRR}, arXiv:1609.07843, 2016.

\bibitem{MnihetalCoRR-13}
Volodymyr Mnih, Koray Kavukcuoglu, David Silver, Alex Graves, Ioannis
  Antonoglou, Daan Wierstra, and Martin Riedmiller.
\newblock Playing {Atari} with deep reinforcement learning.
\newblock {\em CoRR}, arXiv:1312.5602, 2013.

\bibitem{MonperrusACM-17}
Martin Monperrus.
\newblock Automatic software repair: A bibliography.
\newblock {\em {ACM} Computing Surveys}, 51(1):17:1-17:24, 2018.

\bibitem{MoscovitchetalARP-16}
M.~Moscovitch, R.~Cabeza, G.~Winocur, and L.~Nadel.
\newblock Episodic memory and beyond: {T}he hippocampus and neocortex in
  transformation.
\newblock {\em Annual Review of Psychology}, 67:105-134, 2016.

\bibitem{NairetalCoRR-15}
Arun Naira, Praveen Srinivasana, Sam Blackwella, Cagdas Alciceka, Rory Fearona,
  Alessandro~De Mariaa, Vedavyas Panneershelvama, Mustafa Suleymana, Charles
  Beattiea, Stig Petersena, Shane Legga, Volodymyr Mniha, Koray Kavukcuoglua,
  and David Silver.
\newblock Massively parallel methods for deep reinforcement learning.
\newblock {\em CoRR}, arXiv:1507.04296, 2015.

\bibitem{NeelakantanetalICLR-17}
Arvind Neelakantan, Quoc~V. Le, Mart\'{i}n Abadi, Andrew McCallum, and Dario
  Amodei.
\newblock Learning a natural language interface with neural programmer.
\newblock In {\em International Conference on Learning Representations}, volume
  arXiv:1611.08945, 2017.

\bibitem{NielsenNLM-15}
T.~Nielsen, C.~O'Reilly, M.~Carr, G.~Dumel, I.~Godin, E.~Solomonova,
  J.~Lara-Carrasco, C.~Blanchette-Carriere, and T.~Paquette.
\newblock Overnight improvements in two {R}{E}{M} sleep-sensitive tasks are
  associated with both {REM} and {NREM} sleep changes, sleep spindle features,
  and awakenings for dream recall.
\newblock {\em Neurobiology Learning and Memory}, 122:88-97, 2015.

\bibitem{OReillySCIENCE-06}
Randall~C. O'Reilly.
\newblock Biologically based computational models of high-level cognition.
\newblock {\em Science}, 314:91-94, 2006.

\bibitem{OReillyandFrankNC-06}
Randall~C. O'Reilly and Michael~J. Frank.
\newblock Making working memory work: A computational model of learning in the
  prefrontal cortex and basal ganglia.
\newblock {\em Neural Computation}, 18:283-328, 2006.

\bibitem{OReillyetalLEABRA-16}
Randall~C. O'Reilly, Thomas~E. Hazy, and Seth~A. Herd.
\newblock The {Leabra} cognitive architecture: {H}ow to play 20 principles with
  nature and win!
\newblock In Susan E.~F. Chipman, editor, {\em The Oxford Handbook of Cognitive
  Science}, Oxford Handbooks, pages 91-115. Oxford University Press, 2016.

\bibitem{OReillyetalCS-15}
Randall~C. O’Reilly, Rajan Bhattacharyya, Michael~D. Howard, and Nicholas
  Ketz.
\newblock Complementary learning systems.
\newblock {\em Cognitive Science}, 38(6):1229-1248, 2014.

\bibitem{OReillyetalTACO-14}
Randall~C. O’Reilly, Alex~A. Petrov, Jonathan~D. Cohen, Christian~J. Lebiere,
  Seth~A. Herd, and Trent Kriete.
\newblock How limited systematicity emerges: A computational cognitive
  neuroscience approach.
\newblock In Paco Calvo and John Symons, editors, {\em The Architecture of
  Cognition}, pages 191-224. MIT Press, Cambridge, Massachusetts, 2014.

\bibitem{PascanuetalCoRR-17}
Razvan Pascanu, Yujia Li, Oriol Vinyals, Nicolas Heess, Lars Buesing,
  S{\'{e}}bastien Racani{\`{e}}re, David~P. Reichert, Theophane Weber, Daan
  Wierstra, and Peter Battaglia.
\newblock Learning model-based planning from scratch.
\newblock {\em CoRR}, arXiv:1707.06170, 2017.

\bibitem{PenfieldandBoldreyBRAIN-37}
Wilder Penfield and Edwin Boldrey.
\newblock Somatic motor and sensory representation in the cerebral cortex of
  man as studied by electrical stimulation.
\newblock {\em Brain}, 60(4):389-443, 1937.

\bibitem{PenningtonetalEMNLP-14}
Jeffrey Pennington, Richard Socher, and Christopher~D. Manning.
\newblock Glove: Global vectors for word representation.
\newblock In {\em Empirical Methods in Natural Language Processing (EMNLP)},
  pages 1532-1543, 2014.

\bibitem{PiechetalICML-15}
Chris Piech, Jonathan Huang, Andy Nguyen, Mike Phulsuksombati, Mehran Sahami,
  and Leonidas Guibas.
\newblock Learning program embeddings to propagate feedback on student code.
\newblock In {\em Proceedings of the 32nd International Conference on
  International Conference on Machine Learning - Volume 37}, pages 1093-1102,
  2015.

\bibitem{PritzeletalICML-17}
Alexander Pritzel, Benigno Uria, Sriram Srinivasan, Adri{\`a}~Puigdom{\`e}nech
  Badia, Oriol Vinyals, Demis Hassabis, Daan Wierstra, and Charles Blundell.
\newblock Neural episodic control.
\newblock In Doina Precup and Yee~Whye Teh, editors, {\em Proceedings of the
  34th International Conference on Machine Learning}, volume~70 of {\em
  Proceedings of Machine Learning Research}, pages 2827-2836, International
  Convention Centre, Sydney, Australia, 2017. PMLR.

\bibitem{PritzeletalCoRR-17}
Alexander Pritzel, Benigno Uria, Sriram Srinivasan, Adri\`{a}~Puigdom\`{e}nech
  Badia, Oriol Vinyals, Demis Hassabis, Daan Wierstra, and Charles Blundell.
\newblock Neural episodic control.
\newblock {\em CoRR}, arXiv:1703.01988, 2017.

\bibitem{RabinowitzetalCoRR-18}
Neil~C. Rabinowitz, Frank Perbet, H.~Francis Song, Chiyuan Zhang, S.M.~Ali
  Eslami, and Matthew Botvinick.
\newblock Machine theory of mind.
\newblock {\em CoRR}, arXiv:1802.07740, 2018.

\bibitem{ReedandDeFreitasCoRR-15}
Scott~E. Reed and Nando de~Freitas.
\newblock Neural programmer-interpreters.
\newblock {\em CoRR}, arXiv:1511.06279, 2015.

\bibitem{RichandWatersAIM-87}
Charles Rich and Richard~C. Waters.
\newblock The programmer's apprentice: A research overview.
\newblock {\em Computer}, 21(11):10-25, 1988.

\bibitem{RiedmilleretalCoRR-18}
Martin~A. Riedmiller, Roland Hafner, Thomas Lampe, Michael Neunert, Jonas
  Degrave, Tom~Van de~Wiele, Volodymyr Mnih, Nicolas Heess, and Jost~Tobias
  Springenberg.
\newblock Learning by playing - solving sparse reward tasks from scratch.
\newblock {\em CoRR}, arXiv:1802.10567, 2018.

\bibitem{RumelhartetalPDP-86b}
D.~E. Rumelhart, G.~E. Hinton, and J.~L. McClelland.
\newblock A general framework for parallel distributed processing.
\newblock In D.~E. Rumelhart and J.~L. McClelland, editors, {\em Parallel
  Distributed Processing, Volume 1 - Explorations in the Microstructure of
  Cognition: Foundations}, pages 45-76. MIT Press, Cambridge, MA, 1986.

\bibitem{SeeetalACL-17}
Abigail See, Peter~J. Liu, and Christopher~D. Manning.
\newblock Get to the point: Summarization with pointer-generator networks.
\newblock In {\em Proceedings of the 56th Annual Meeting of the Association for
  Computational Linguistics}, volume arXiv:1704.04368, 2017.

\bibitem{ShinetalICLR-18b}
Richard Shin, Illia Polosukhin, and Dawn Song.
\newblock Towards specification-directed program repair.
\newblock In {\em Submitted to International Conference on Learning
  Representations}, 2018.

\bibitem{SilveretalICML-17}
David Silver, Hado van Hasselt, Matteo Hessel, Tom Schaul, Arthur Guez, Tim
  Harley, Gabriel Dulac{-}Arnold, David~P. Reichert, Neil~C. Rabinowitz,
  Andr{\'{e}} Barreto, and Thomas Degris.
\newblock The predictron: End-to-end learning and planning.
\newblock {\em Proceedings of the 34th International Conference on Machine
  Learning}, 2017.

\bibitem{SmithandWatermanJMB-81}
T.~F. Smith and M.~S. Waterman.
\newblock Identification of common molecular subsequences.
\newblock {\em Journal of Molecular Biology}, 147(1):195-197, 1981.

\bibitem{SprechmannetalICLR-18}
Pablo Sprechmann, Siddhant~M. Jayakumar, Jack~W. Rae, Alexander Pritzel,
  Adria~Puigdomenech Badia, Benigno Uria, Oriol Vinyals, Demis Hassabis, Razvan
  Pascanu, and Charles Blundell.
\newblock Memory-based parameter adaptation.
\newblock {\em International Conference on Learning Representations}, 2018.

\bibitem{SukhbaataretalCoRR-15}
Sainbayar Sukhbaatar, Arthur Szlam, Jason Weston, and Rob Fergus.
\newblock Weakly supervised memory networks.
\newblock {\em CoRR}, arXiv:1503.08895, 2015.

\bibitem{SutskeverandHintonNIPS-09}
Ilya Sutskever and Geoffrey~E Hinton.
\newblock Using matrices to model symbolic relationship.
\newblock In D.~Koller, D.~Schuurmans, Y.~Bengio, and L.~Bottou, editors, {\em
  Advances in Neural Information Processing Systems 21}, pages 1593-1600.
  Curran Associates, Inc., 2009.

\bibitem{SuttonetalAIJ-99}
Richard~S. Sutton, Doina Precup, and Satinder Singh.
\newblock Between {MDP}s and semi-{MDP}s: A framework for temporal abstraction
  in reinforcement learning.
\newblock {\em Artificial. Intelligence}, 112(1-2):181-211, 1999.

\bibitem{TamaretalNIPS-16}
Aviv Tamar, Yi~Wu, Garrett Thomas, Sergey Levine, and Pieter Abbeel.
\newblock Value iteration networks.
\newblock In {\em Proceedings of the 30th International Conference on Neural
  Information Processing Systems}. Curran Associates Inc., 2016.

\bibitem{TeylerandRudyHIPPOCAMPUS-07}
T.~J. Teyler and J.~W. Rudy.
\newblock The hippocampal indexing theory and episodic memory: updating the
  index.
\newblock {\em Hippocampus}, 17(12):1158-1169, 2007.

\bibitem{TishbyandZaslavskyCoRR-15}
Naftali Tishby and Noga Zaslavsky.
\newblock Deep learning and the information bottleneck principle.
\newblock {\em CoRR}, arXiv:1503.02406, 2015.

\bibitem{Tulving1972}
E.~Tulving, W.~Donaldson, and G.H. Bower.
\newblock {\em Organization of memory}.
\newblock Academic Press, 1972.

\bibitem{VinyalsetalNIPS-15}
Oriol Vinyals, Meire Fortunato, and Navdeep Jaitly.
\newblock Pointer networks.
\newblock In C.~Cortes, N.~D. Lawrence, D.~D. Lee, M.~Sugiyama, and R.~Garnett,
  editors, {\em Advances in Neural Information Processing Systems 28}, pages
  2692-2700. Curran Associates, Inc., 2015.

\bibitem{VinyalsandLeICML-15}
Oriol Vinyals and Quoc~V. Le.
\newblock A neural conversational model.
\newblock In {\em ICML Deep Learning Workshop}, 2015.

\bibitem{WangandJiangICLR-17}
Shuohang Wand and Jing Jiang.
\newblock Machine comprehension using match-{LSTM} and answer pointer.
\newblock In {\em International Conference on Learning Representations}, volume
  arXiv:1608.07905, 2017.

\bibitem{WangetalNATURE-NEUROSCIENCE-18}
Jane~X Wang, Zeb Kurth-Nelson, Dharshan Kumaran, Dhruva Tirumala, Hubert Soyer,
  Joel~Z Leibo, Demis Hassabis, and Matthew Botvinick.
\newblock Prefrontal cortex as a meta-reinforcement learning system.
\newblock {\em Nature Neuroscience}, 21:860-868, 2018.

\bibitem{WangetalCoRR-17}
Ke~Wang, Rishabh Singh, and Zhendong Su.
\newblock Dynamic neural program embedding for program repair.
\newblock {\em CoRR}, arXiv:1711.07163, 2017.

\bibitem{WayneetalCoRR-18}
Greg Wayne, Chia-Chun Hung, David Amos, Mehdi Mirza, Arun Ahuja, Agnieszka
  Grabska-Barwinska, Jack Rae, Piotr Mirowski, Joel~Z. Leibo, Adam Santoro,
  Mevlana Gemici, Malcolm Reynolds, Tim Harley, Josh Abramson, Shakir Mohamed,
  Danilo Rezende, David Saxton, Adam Cain, Chloe Hillier, David Silver, Koray
  Kavukcuoglu, Matt Botvinick, Demis Hassabis, and Timothy Lillicrap.
\newblock Unsupervised predictive memory in a goal-directed agent.
\newblock {\em CoRR}, arXiv:1803.10760, 2018.

\bibitem{WeberetalCoRR-17}
Theophane Weber, S{\'{e}}bastien Racani{\`{e}}re, David~P. Reichert, Lars
  Buesing, Arthur Guez, Danilo~Jimenez Rezende, Adri{\`{a}}~Puigdom{\`{e}}nech
  Badia, Oriol Vinyals, Nicolas Heess, Yujia Li, Razvan Pascanu, Peter
  Battaglia, David Silver, and Daan Wierstra.
\newblock Imagination-augmented agents for deep reinforcement learning.
\newblock {\em CoRR}, arXiv:1707.06203, 2017.

\bibitem{WestonetalCoRR-14}
Jason Weston, Sumit Chopra, and Antoine Bordes.
\newblock Memory networks.
\newblock {\em CoRR}, arXiv:1410.3916, 2014.

\bibitem{CS379C_Final_Project_Wong-18}
Catherine Wong.
\newblock Reconstructive memory for abstract selective recall.
\newblock Representative Final Project Archive for Stanford CS379C in Spring
  2018, 2018.

\bibitem{XuetalCoRR-17}
Xiaojun Xu, Chang Liu, Qian Feng, Heng Yin, Le~Song, and Dawn Song.
\newblock Neural network-based graph embedding for cross-platform binary code
  similarity detection.
\newblock {\em CoRR}, arXiv:1708.06525, 2017.

\bibitem{YinandNeubigACL-17}
Pengcheng Yin and Graham Neubig.
\newblock A syntactic neural model for general-purpose code generation.
\newblock In {\em The 55th Annual Meeting of the Association for Computational
  Linguistics (ACL)}, Vancouver, Canada, 2017.

\bibitem{ZeileretalCVPR-10}
Matthew~D. Zeiler, Dilip Kirshnan, Graham~W. Taylor, and Rob Fergus.
\newblock Deconvolutional networks.
\newblock In {\em {IEEE} International Conference on Computer Vision and
  Pattern Recognition}, pages 2528-2535, 2010.

\bibitem{ZeileretalICCV-11}
M.D. Zeiler, G.W. Taylor, and R.~Fergus.
\newblock Adaptive deconvolutional networks for mid and high level feature
  learning.
\newblock In {\em {IEEE} International Conference on Computer Vision}, pages
  2018-2025, 2011.

\end{thebibliography}
